\documentclass[universe,review,accept,moreauthors,pdftex]{mdpi} 

\setitemize{parsep=6pt,itemsep=0pt,leftmargin=*,labelsep=5.5mm}
\setenumerate{parsep=6pt,itemsep=0pt,leftmargin=*,labelsep=5.5mm}
\setlist[description]{itemsep=0mm}

\firstpage{1} 
\pubvolume{5}
\issuenum{5}
\articlenumber{110}
\pubyear{2019}
\copyrightyear{2019}
\history{Received: 25 February 2019; Accepted: 5 May 2019; Published: 9 May 2019}
\updates{yes}

\usepackage[utf8]{inputenc}

\usepackage{amsmath,amssymb}

\Title{Induced Gravitational Collapse, Binary-Driven Hypernovae, Long
Gramma-ray Bursts and Their Connection with Short Gamma-ray Bursts}

\Author{J. A. Rueda $^{1,2,3,}$*, R. Ruffini $^{1,2,4}$ and Y. Wang $^{1,2}$}
\address{%
$^{1}$ \quad ICRANet, Piazza della Repubblica 10, 65122~Pescara, Italy; ruffini@icra.it (R.R.); wangyu@me.com (Y.W.)\\
$^{2}$ \quad ICRA, Dipartimento di Fisica, Universit\`a di Roma Sapienza, Piazzale Aldo Moro 5, 00185 Rome,  Italy\\
$^{3}$ \quad INAF, Istituto de Astrofisica e Planetologia Spaziali, Via Fosso del Cavaliere 100, 00133 Rome, Italy\\
$^{4}$ \quad INAF, Viale del Parco Mellini 84, 00136 Rome, Italy}

\corres{Correspondence: jorge.rueda@icra.it}

\abstract{There is increasing observational evidence that short and long Gamma-ray bursts (GRBs) originate in different subclasses, each one with specific energy release, spectra, duration, etc, and all of them with binary progenitors. The binary components involve carbon-oxygen cores (CO$_\textrm{core}$), neutron stars (NSs), black holes (BHs), and white dwarfs (WDs). We review here the salient features of the specific class of binary-driven hypernovae (BdHNe) within the induced gravitational collapse (IGC) scenario for the explanation of the long GRBs. The progenitor is a~CO$_\textrm{core}$-NS binary. The~supernova (SN) explosion of the CO$_\textrm{core}$, producing at its center a~new NS ($\nu$NS), triggers onto the NS companion a~hypercritical, i.e., highly super-Eddington accretion process, accompanied by a~copious emission of neutrinos. By accretion the NS can become either a~more massive NS or reach the critical mass for gravitational collapse with consequent formation of a~BH. We summarize the results on this topic from the first analytic estimates in 2012 all the way up to the most recent three-dimensional (3D) smoothed-particle-hydrodynamics (SPH) numerical simulations in 2018. Thanks~to these results it is by now clear that long GRBs are richer and more complex systems than thought before. {The~SN explosion and its  hypercritical accretion onto the NS explain the X-ray precursor. The~feedback of the NS accretion, the~NS collapse and the BH formation produce asymmetries in the SN ejecta, implying the necessity of a~3D analysis for GRBs. The~newborn BH, the surrounding matter and the magnetic field inherited from the NS, comprises the \emph{inner engine} from which the GRB electron-positron ($e^+e^-$) plasma and the high-energy emission are initiated. The~impact of the $e^+e^-$ on the asymmetric ejecta transforms the SN into a~hypernova (HN). The~dynamics of the plasma in the asymmetric ejecta leads to signatures depending on the viewing angle. This~explains the ultrarelativistic prompt emission in the MeV domain and the mildly-relativistic flares in the early afterglow in the X-ray domain. The~feedback of the $\nu$NS pulsar-like emission on the HN explains the X-ray late afterglow and its power-law regime. All~of the above} is in contrast with a~simple GRB model attempting to explain the entire GRB with the kinetic energy of an~ultrarelativistic jet extending through all of the above GRB phases, as traditionally proposed in the ``collapsar-fireball'' model. In~addition, BdHNe in their different flavors lead to $\nu$NS-NS or $\nu$NS-BH binaries. The~gravitational wave emission drives these binaries to merge producing short GRBs. It~is thus established a~previously unthought interconnection between long and short GRBs and their occurrence rates. This needs to be accounted for in the cosmological evolution of binaries within population synthesis models for the formation of compact-object binaries.}


\begin{document}

\section{Introduction}

%%%%%%%%%%%%%%%%%%%%%%%%%%%%%%%%%%%%%%%%%%%%%%%%%%%%%%%%%%%%
\subsection{The Quest for the Binary Nature of GRB Progenitors}
%%%%%%%%%%%%%%%%%%%%%%%%%%%%%%%%%%%%%%%%%%%%%%%%%%%%%%%%%%%%

We first recall that GRBs have been traditionally classified by a~phenomenological division based on the duration of the time-interval in which the 90\% of the total isotropic energy in Gamma-rays is emitted, the $T_{90}$. Long GRBs are those with $T_{90}>2$~s and short GRBs the sources with $T_{90}<2$~s~\cite{1981Ap&SS..80....3M,1992grbo.book..161K,1992AIPC..265..304D,1993ApJ...413L.101K,1998ApJ...497L..21T}.

In the case of short bursts, rapid consensus was reached in the scientific community that they could be the product of mergers of NS-NS and/or NS-BH binaries (see e.g., the pioneering works~\cite{1986ApJ...308L..47G,1986ApJ...308L..43P,1989Natur.340..126E,1991ApJ...379L..17N}). We shall return on this issue below by entering into the description of their properties and also to introduce additional mergers of compact-star object binaries leading to short bursts.

For long bursts, possibly the most compelling evidence of the necessity of a~binary progenitor comes from the systematic and spectroscopic analysis of the GRBs associated with SNe, the so-called GRB-SNe, started with the pioneering discovery of the spatial and temporal concomitance of GRB 980425~\cite{2000ApJ...536..778P} and SN 1998bw~\cite{1998Natur.395..670G}. Soon after, many associations of other nearby GRBs with type Ib/c SNe were evidenced (see e.g.,~\cite{2011IJMPD..20.1745D,2017AdAst2017E...5C}). 

There are models in the literature attempting an~explanation of both the SN and the GRB within the same astrophysical system. For instance, GRBs have been assumed to originate from a~violent SN from the collapse of a~massive and fast rotating star, a~``collapsar''~\cite{2006ARA&A..44..507W}. A very high rotating rate of the star is needed to produce a~collimated, jet emission. {This traditional picture adopts for the GRB dynamics the ``fireball'' model based on the existence of a~single ultrarelativistic collimated jet~\cite{1976PhFl...19.1130B,1990ApJ...365L..55S,1993ApJ...415..181M,1993MNRAS.263..861P,1994ApJ...424L.131M}. There is a~vast literature devoted to this ``traditional" approach and we refer the reader to it for additional details (see, e.g.,~\cite{1999PhR...314..575P,2004RvMP...76.1143P,2002ARA26A..40..137M,2006RPPh...69.2259M,2014ARA&A..52...43B,2015PhR...561....1K}, and references therein)}. 

{Nevertheless, it is worth to mention here some of the most important drawbacks of the aforementioned``traditional'' approach and which has motivated the introduction of an~alternative model, based on a~binary progenitor, for the explanation of long GRBs:}
\begin{itemize}
\item
{
SNe Ic as the ones associated with GRBs lack hydrogen and helium in their spectra. It has been recognized that they most likely originate in helium stars, CO$_{\rm core}$, or Wolf-Rayet stars, that~have lost their outermost layers (see e.g.,~\cite{2011MNRAS.415..773S}, and references therein). The~pre-SN star, very~likely, does~not follow a~single-star evolution but it belongs to a~tight binary with a~compact star companion (e.g., a~NS). The~compact star strips off the pre-SN star outermost layers via binary interactions such as mass-transfer and tidal effects} (see e.g.,~\cite{1988PhR...163...13N,1994ApJ...437L.115I,2007PASP..119.1211F,2010ApJ...725..940Y,2011MNRAS.415..773S}).

\item
Denoting the beaming angle by $\theta_j$, to an~observed isotropic energy $E_{\rm iso}$ it would correspond to a~reduced intrinsic source energy released $E_{s}=f_b E_{\rm iso}<E_{\rm iso}$, where $f_b =(1-\cos\theta_j)\sim \theta^2_j/2<1$. Extremely small beaming factors $f_b\sim 1/500$ (i.e., $\theta_j \sim 1^\circ$) are inferred to reduce the observed energetics of $E_{\rm iso}\sim 10^{54}$~erg to the expected energy release by such a~scenario $\sim$10$^{51}$~erg~\cite{2001ApJ...562L..55F}. However, the existence of such extremely narrow beaming angles have never been observationally corroborated~\cite{2006NCimB.121.1171C,2007ApJ...657..359S,2009cfdd.confE..23B}. 
\item
An additional drawback of this scenario is that it implies a~dense and strong wind-like circumburst medium (CBM) in contrast with the one observed in most GRBs (see e.g.,~\cite{2012A&A...543A..10I}). Indeed, the average CBM density inferred from GRB afterglows is of the order of 1 baryon per cubic centimeter~\cite{2011IJMPD..20.1797R}. The baryonic matter component in the GRB process is represented by the so-called baryon load~\cite{2000A&A...359..855R}. The GRB $e^+e^-$ plasma should engulf a~limited amount of baryons in order to be able to expand at ultrarelativistic velocities with Lorentz factors $\Gamma\gtrsim 100$ as requested by the observed non-thermal component in the prompt Gamma-ray emission spectrum~\cite{1990ApJ...365L..55S,1993MNRAS.263..861P,1993ApJ...415..181M}. The amount of baryonic mass $M_B$ is thus limited by the prompt emission to a~maximum value of the baryon-load parameter, $B=M_B c^2/E_{e^+e-}\lesssim 10^{-2}$, where $E_{e^+e-}$ is the total energy of the $e^+e-$ plasma~\cite{2000A&A...359..855R}. 
\item
{
GRBs and SNe have markedly different energetics. SNe emit energies in the range $10^{49}$--$10^{51}$~erg, while GRBs emit in the range $10^{49}$--$10^{54}$~erg. Thus, the origin of GRB energetics point to the gravitational collapse to a~stellar-mass BH. The SN origin points to evolutionary stages of a~massive star leading to a~NS or to a~complete disrupting explosion, but not to a~BH. The direct formation of a~BH in a~core-collapse SN is currently ruled out by the observed masses of pre-SN progenitors, $\lesssim$18~$M_\odot$~\cite{2015PASA...32...16S}. It is theoretically known that massive stars with such a~relatively low mass do not lead to a~direct collapse to a~BH (see~\cite{2009ARA&A..47...63S,2015PASA...32...16S} for details)
}.
\item
{
It was recently shown in~\cite{2018ApJ...852...53R} that the observed thermal emission in the X-ray flares present in the early (rest-frame time $t\sim 10^2$~s) afterglow implies an~emitter of size $\sim$10$^{12}$~cm expanding at mildly-relativistic velocity, e.g.,~$\Gamma\lesssim 4$. This is clearly in contrast with the ``collapsar-fireball'' scenario in which there is an~ultrarelativistic emitter (the jet) with $\Gamma \sim 10^2$--$10^3$ extending from the prompt emission all the way to the afterglow.
}
\end{itemize}

Therefore, it seems most unlikely that the GRB and the SN can originate from the same single-star progenitor. Following this order of ideas, it was introduced for the explanation of the spatial and temporal coincidence of the two phenomena the concept of  \emph{induced gravitational collapse} (IGC)~\cite{2001ApJ...555L.117R,2008mgm..conf..368R}. Two scenarios for the GRB-SN connection have been addressed: Ruffini et al.~\cite{2001ApJ...555L.117R} considered that the GRB was the trigger of the SN. However, for this scenario to happen it was shown that the companion star had to be in a~very fine-tuned phase of its stellar evolution~\cite{2001ApJ...555L.117R}. \mbox{Ruffini et al.~\cite{2008mgm..conf..368R}} proposed an~alternative scenario in a~compact binary: the explosion of a~Ib/c SN triggering an~accretion process onto a~NS companion. The NS, reaching the critical mass value, gravitationally collapses leading to the formation of a~BH. The formation of the BH consequently leads to the emission of the GRB. Much more about this binary scenario has been discovered since its initial proposal; its theoretical studies and the search for its observational verification have led to the formulation of a~much rich phenomenology which will be the main subject of this article. 

Therefore, both short and long GRBs appear to be produced by binary systems, well in line with the expectation that most massive stars belong to binary systems {(see, e.g.,~\cite{2012Sci...337..444S,2014ARA&A..52..487S}, and references therein)}. The increasing amount and quality of the multiwavelength data of GRBs have revealed the richness of the GRB phenomenon which, in a~few seconds, spans different regimes from X-ray precursors to the Gamma-rays of the prompt emission, to the optical and X-rays of the early and late afterglow, to the optical emission of the associated SNe and, last but not least, the presence or absence of high-energy GeV emission. This, in addition to the multiyear effort of reaching a~comprehensive theoretical interpretation of such regimes, have lead to the conclusion that GRBs separate into subclasses, each with specific energy release, spectra, duration, among other properties and, indeed, all with binary progenitors~\cite{2016ApJ...832..136R,2017IJMPD..2630016R,2018ApJ...859...30R,2018arXiv180305476R,2019ApJ...874...39W}.

%%%%%%%%%%%%%%%%%%%%%%%%%%%%%%%%%%%%%%%%%%%%%%%%%%%%%%%%%%%%
\subsection{GRB Subclasses}
%%%%%%%%%%%%%%%%%%%%%%%%%%%%%%%%%%%%%%%%%%%%%%%%%%%%%%%%%%%%

Up to 2017 we had introduced seven GRB subclasses summarized in Table~\ref{tab:rates}. In addition, we have recently introduced in~\cite{2018JCAP...10..006R,2019JCAP...03..044R} the possibility of a~further GRB subclass produced by WD-WD mergers. We now give a~brief description of all the GRB subclasses identified. In~\cite{2019ApJ...874...39W} we have renominated the GRB subclasses introduced in~\cite{2016ApJ...832..136R} and in~\cite{2018JCAP...10..006R,2019JCAP...03..044R}, and inserted them into two groups: binary-driven hypernovae (BdHNe) and compact-object binary mergers. Below we report both the old and the new names to facilitate the reader when consulting our works prior to~\cite{2019ApJ...874...39W}.

\begin{enumerate}[align=parleft,leftmargin=*,labelsep=3mm]
\item[i.]
\textbf{X-ray flashes (XRFs)}. 
These systems have CO$_{\rm core}$-NS binary progenitors in which the NS companion does not reach the critical mass for gravitational collapse~\cite{2016ApJ...833..107B,2015ApJ...812..100B}. In the SN explosion, the binary might or might not be disrupted depending on the mass loss and/or the kick imparted~\cite{2014LRR....17....3P}. Thus XRFs lead either to two NSs {ejected by the disruption}, or to binaries composed of a~newly-formed $\sim$1.4--1.5~$M_\odot$ NS (hereafter $\nu$NS) born at the center of the SN, and a~massive NS (MNS) which accreted matter from the SN ejecta. Some observational properties are: Gamma-ray isotropic energy $E_{\rm iso}\lesssim 10^{52}$~erg, rest-frame spectral peak energy $E_{p,i}\lesssim 200$~keV and a~local observed rate of $\rho_{\rm XRF}=100^{+45}_{-34}$~Gpc$^{-3}$~yr$^{-1}$~\cite{2016ApJ...832..136R}. We refer the reader to Table~\ref{tab:rates} and~\cite{2016ApJ...832..136R,2018ApJ...859...30R} for further details on this class. In~\cite{2019ApJ...874...39W}, this class has been divided into BdHN type II, the sources with $10^{50}\lesssim E_{\rm iso}\lesssim 10^{52}$~erg, and BdHN type III, the sources with $10^{48}\lesssim E_{\rm iso}\lesssim 10^{50}$~erg.
\item[ii.]
\textbf{Binary-driven hypernovae (BdHNe)}. Originate in compact CO$_{\rm core}$-NS binaries where the accretion onto the NS becomes high enough to bring it to the point of gravitational collapse, hence forming a~BH. We showed that most of these binaries survive to the SN explosion owing to the short orbital periods ($P\sim 5$~min) for which the mass loss cannot be considered as instantaneous, allowing the binary to keep bound even if more than half of the total binary mass is lost~\cite{2015PhRvL.115w1102F}. Therefore, BdHNe produce $\nu$NS-BH binaries. Some observational properties are: $E_{\rm iso}\gtrsim10^{52}$~erg, $E_{p,i}\gtrsim200$~keV and a~local observed rate of $\rho_{\rm BdHN}=0.77^{+0.09}_{-0.08}$~Gpc$^{-3}$~yr$^{-1}$~\cite{2016ApJ...832..136R}. We refer the reader to Table~\ref{tab:rates} and~\cite{2016ApJ...832..136R,2018ApJ...859...30R} for further details on this class. In~\cite{2019ApJ...874...39W} this class has been renominated as BdHN type I.

\item[iii.]
\textbf{BH-SN}. These systems originate in CO$_{\rm core}$ (or Helium or Wolf-Rayet star)-BH binaries, hence the hypercritical accretion of the SN explosion of the CO$_{\rm core}$ occurs onto a~BH previously formed in the evolution path of the binary. They might be the late evolutionary stages of X-ray binaries such as Cyg X-1~\cite{1978pans.proc.....G,2011ApJ...742L...2B}, or microquasars~\cite{1998Natur.392..673M}.~Alternatively, they can form following the evolutionary scenario XI in~\cite{1999ApJ...526..152F}. If the binary survives to the SN explosion BH-SNe produce $\nu$NS-BH, or~BH-BH binaries when the central remnant of the SN explosion collapses directly to a~BH (see,~although,~\mbox{\cite{2009ARA&A..47...63S,2015PASA...32...16S}}). Some observational properties are: $E_{\rm iso}\gtrsim10^{54}$~erg,  $E_{p,i}\gtrsim2$~MeV and an~upper limit to their rate is $\rho_{\rm BH-SN}\lesssim \rho_{\rm BdHN} = 0.77^{+0.09}_{-0.08}$~Gpc$^{-3}$~yr$^{-1}$, namely the estimated observed rate of BdHNe type I which by definition covers systems with the above $E_{\rm iso}$ and $E_{p,i}$ range~\cite{2016ApJ...832..136R}. We~refer the reader to Table~\ref{tab:rates} and~\cite{2016ApJ...832..136R,2018ApJ...859...30R} for further details on this class. In~\cite{2019ApJ...874...39W} this class has been renominated as BdHN type IV. 
\end{enumerate}

\begin{table}[H]
\caption{Summary of the Gamma-ray bursts (GRB) subclasses. This table is an~extended version of the one presented in~\cite{2019ApJ...874...39W} with the addition of a~column showing the local density rate, and it also updates the one in~\cite{2016ApJ...832..136R,2018ApJ...859...30R}. We unify here all the GRB subclasses under two general names, BdHNe and BMs. Two new GRB subclasses are introduced; BdHN Type III and BM Type IV. In~addition to the subclass name in ``Class'' column and ``Type'' column, as well as the previous names in ``Previous Alias'' column, we report the number of GRBs with known redshift identified in each subclass updated by the end of 2016 in ``number'' column (the value in a~bracket indicates the lower limit). We recall as well the ``in-state'' representing the progenitors and the ``out-state'' representing the outcomes, as well as the the peak energy of the prompt emission, $ E_{\rm p,i}$, the isotropic Gamma-ray energy, $E_{\rm iso}$~defined in the $1$~keV to $10$~MeV energy range, the isotropic emission of ultra-high energy photons, $E_{\rm iso,Gev}$, defined in the $0.1$--$100$ GeV energy range, and the local observed rate $\rho_{\rm GRB}$~\cite{2016ApJ...832..136R}. We adopt as definition of kilonova a~phenomenon more energetic than a~nova (about 1000 times). A kilonova can be an~infrared-optical counterpart of a~NS-NS merger. In that case the transient is powered by the energy release from the decay of r-process heavy nuclei processed in the merger ejecta~\cite{1998ApJ...507L..59L,2010MNRAS.406.2650M,2013Natur.500..547T,2013ApJ...774L..23B}. FB-KN stands for fallback-powered kilonova~\cite{2018JCAP...10..006R,2019JCAP...03..044R}: a~WD-WD merger can emit an~infrared-optical transient, peaking at $\sim$5~day post-merger, with the ejecta powered by accretion of fallback matter onto the newborn WD formed in the merger. {The density rate of the GRB subclasses BdHN III (HN) and BM IV (FB-KN) have not yet been estimated.}}
\label{tab:rates}
\centering
%\tablesize{\scriptsize}
\scalebox{0.72}[0.72]{\begin{tabular}{cccccccccc}
\toprule
\textbf{Class} &   \textbf{Type}  & \textbf{Previous  Alias} & \textbf{Number} & \textbf{\emph{In-State}}  & \textbf{\emph{Out-State}} & \boldmath$E_{\rm p, i}$ \textbf{(MeV)}&  \boldmath$E_{\rm iso} \textbf{(erg)}$  &  \boldmath$E_{\rm iso,Gev}$  \textbf{(erg)}& \boldmath$\rho_{\rm GRB}$  \textbf{(Gpc}\boldmath$^{-3}$ \textbf{yr}$^{-1}$\textbf{)}\\
%		
%& &  & & & &  &  &   & 	\\	
\midrule
Binary-driven & I  & BdHN  & $329$ &CO$_{\rm core}$-NS  & $\nu$NS-BH & $\sim$0.2--$2$ &  $\sim$10$^{52}$--$10^{54}$ &    $\gtrsim$10$^{52}$ & $0.77^{+0.09}_{-0.08}$\\
hypernova & II & XRF & $(30)$ &CO$_{\rm core}$-NS    & $\nu$NS-NS & $\sim$0.01--$0.2$  &  $\sim$10$^{50}$--$10^{52}$ &    $-$ &  $100^{+45}_{-34}$\\

(BdHN) & III  & HN & $(19)$ & CO$_{\rm core}$-NS  & $\nu$NS-NS & $\sim$0.01 &  $\sim$10$^{48}$--$10^{50}$ &  $-$ &  $-$ \\
& IV   & BH-SN & $5$ & CO$_{\rm core}$-BH  & $\nu$NS-BH & $\gtrsim$2 &  $>$10$^{54}$ &   $\gtrsim$10$^{53}$ &  $\lesssim$0.77$^{+0.09}_{-0.08}$ \\
\midrule
& I & S-GRF & $18$ &NS-NS & MNS & $\sim$0.2--$2$ &  $\sim$10$^{49}$--$10^{52}$  &  $-$ &  $3.6^{+1.4}_{-1.0}$\\
Binary & II  & S-GRB  & $6$ &NS-NS & BH & $\sim$2--$8$ &  $\sim$10$^{52}$--$10^{53}$ &   $\gtrsim$10$^{52}$ &  $\left(1.9^{+1.8}_{-1.1}\right)\times10^{-3}$\\
Merger& III  & GRF  & $(1)$ &NS-WD & MNS & $\sim$0.2--$2$ &  $\sim$10$^{49}$--$10^{52}$ & $-$ & $1.02^{+0.71}_{-0.46}$\\
(BM)& IV  & FB-KN & $(1)$ &WD-WD & NS/MWD & $<$0.2 &  $<$10$^{51}$  & $-$ & $-$\\
& V   & U-GRB & $(0)$ &NS-BH & BH & $\gtrsim$2 &  $>$10$^{52}$ & $-$ & $\approx$0.77$^{+0.09}_{-0.08}$\\
\bottomrule
\end{tabular}}
\end{table}

We proceed with the short bursts which are amply thought to originate from compact-object binary mergers (BMs). First, we discuss the traditionally proposed BMs namely NS-NS and/or NS-BH mergers~\cite{1986ApJ...308L..47G,1986ApJ...308L..43P,1989Natur.340..126E,1991ApJ...379L..17N,1997ApJ...482L..29M,2003MNRAS.345.1077R,2004ApJ...608L...5L,2014ARA&A..52...43B}. These BMs can be separated into three subclasses~\cite{2015PhRvL.115w1102F,2015ApJ...808..190R,2016ApJ...832..136R}:
\begin{enumerate}[align=parleft,leftmargin=*,labelsep=3mm]
\item[iv.]
\textbf{Short Gamma-ray flashes (S-GRFs)}. They are produced by NS-NS mergers leading to a~MNS, namely when the merged core does not reach the critical mass of a~NS. Some observational properties are: $E_{\rm iso}\lesssim10^{52}$~erg, $E_{p,i}\lesssim2$~MeV and a~local observed rate of \mbox{$\rho_{\rm S-GRF}=3.6^{+1.4}_{-1.0}$~Gpc$^{-3}$~yr$^{-1}$~\cite{2016ApJ...832..136R}}. % yr is changed to be year, please confirm.
We refer the reader to Table~\ref{tab:rates} and~\cite{2016ApJ...832..136R,2018ApJ...859...30R} for further details on this class. In~\cite{2019ApJ...874...39W} this class has been renominated as BM type I. 
\item[v.]
\textbf{Authentic short GRBs (S-GRBs)}. They are produced by NS-NS mergers leading to a~BH, namely when the merged core reaches the critical mass of a~NS, hence it forms a~BH as a~central \mbox{remnant~\cite{2016ApJ...831..178R,2015ApJ...808..190R,2013ApJ...763..125M}}. Some observational properties are: $E_{\rm iso}\gtrsim 10^{52}$~erg,  $E_{p,i}\gtrsim2$~MeV and a~local observed rate of $\rho_{\rm S-GRB}=\left(1.9^{+1.8}_{-1.1}\right)\times10^{-3}$~Gpc$^{-3}$~yr$^{-1}$~\cite{2016ApJ...832..136R}. We refer the reader to Table~\ref{tab:rates} and~\cite{2016ApJ...832..136R,2018ApJ...859...30R} for further details on this class. In~\cite{2019ApJ...874...39W} this class has been renominated as BM type II.
\item[vi.]
\textbf{Ultra-short GRBs (U-GRBs)}. This is a~theoretical GRB subclass subjected for observational verification. U-GRBs are expected to be produced by $\nu$NS-BH mergers whose binary progenitors can be the outcome of BdHNe type I (see II above) or of BdHNe type IV (BH-SN; see III above). The following observational properties are expected: $E_{\rm iso}\gtrsim 10^{52}$~erg, $E_{p,i}\gtrsim2$~MeV and a~local observed rate similar to the one of BdHNe type I since we have shown that most of them are expected to remain bound~\cite{2015PhRvL.115w1102F}, i.e., $\rho_{\rm U-GRB} \approx \rho_{\rm BdHN} = 0.77^{+0.09}_{-0.08}$~Gpc$^{-3}$~yr$^{-1}$~\cite{2016ApJ...832..136R}. We refer the reader to Table~\ref{tab:rates} and~\cite{2016ApJ...832..136R,2018ApJ...859...30R} for further details on this class. In~\cite{2019ApJ...874...39W} this class has been renominated as BM type V.

\end{enumerate}

Besides the existence of the above three subclasses of long bursts and three subclasses of short bursts in which the presence of NSs plays a~fundamental role, there are two subclasses of bursts in which there is at least a~WD component.

\begin{enumerate}[align=parleft,leftmargin=*,labelsep=4.5mm]
\item[vii.]
\textbf{Gamma-ray flashes (GRFs)}. These sources show an~extended and softer emission, i.e., they have hybrid properties between long and short bursts and have no associated SNe~\cite{2006Natur.444.1050D}. It~has been proposed that they are produced by NS-WD mergers~\cite{2016ApJ...832..136R}. These binaries are expected to be very numerous~\cite{2015ApJ...812...63C} and a~variety of evolutionary scenarios for their formation have been proposed~\cite{1992ApJ...400..175B,1999ApJ...520..650F,2000ApJ...530L..93T,2014MNRAS.437.1485L}. GRFs form a~MNS and not a~BH~\cite{2016ApJ...832..136R}. Some observational properties are:  \mbox{$10^{51}\lesssim E_{\rm iso}\lesssim 10^{52}$~erg}, $0.2 \lesssim E_{p,i}\lesssim 2$~MeV and a~local observed rate of $\rho_{\rm GRF}=1.02^{+0.71}_{-0.46}$~Gpc$^{-3}$~yr$^{-1}$~\cite{2016ApJ...832..136R}. It is worth noting that this rate is low with respect to the one expected from the current number of known NS-WD in the Galaxy~\cite{2015ApJ...812...63C}. From the GRB observations only one NS-WD merger has been identified (GRB 060614~\cite{2009A&A...498..501C}). This implies that most NS-WD mergers are probably under the threshold of current X and Gamma-ray instruments. We refer the reader to Table~\ref{tab:rates} and~\cite{2016ApJ...832..136R,2018ApJ...859...30R} for further details on this class. In~\cite{2019ApJ...874...39W} this class has been renominated as BM type III.
\item[viii.]
\textbf{Fallback kilonovae (FB-KNe)}. This is a~recently introduced GRB subclass having as progenitors WD-WD mergers~\cite{2018JCAP...10..006R,2019JCAP...03..044R}. The WD-WD mergers of interest are those that do not produce type Ia SNe but that lead to a~massive ($M\sim 1~M_\odot$), fast rotating ($P\sim 1$--$10$~s), highly-magnetized ($B\sim 10^9$--$10^{10}$~G) WD. Some observational properties are:  $E_{\rm iso}\lesssim 10^{51}$~erg, $E_{p,i}\lesssim 2$~MeV and a~local observed rate $\rho_{\rm FB-KN} = (3.7$--$6.7)\times 10^5$~Gpc$^{-3}$~yr$^{-1}$~\cite{2018JCAP...10..006R,2019JCAP...03..044R,2017MNRAS.467.1414M,2018MNRAS.476.2584M}. The coined name FB-KN is due to the fact that they are expected to produce an~infrared-optical transient by the cooling of the ejecta expelled in the dynamical phase of the merger and heated up by fallback accretion onto the newly-formed massive WD.

\end{enumerate}

The density rates for all GRB subclasses have been estimated assuming no beaming~\cite{2016ApJ...832..136R,2018ApJ...859...30R,2018JCAP...10..006R,2019JCAP...03..044R}. {The GRB density rates have been analyzed in~\cite{2016ApJ...832..136R} following the method suggested in~\cite{2015ApJ...812...33S}.}

%%%%%%%%%%%%%%%%%%%%%%%%%%%%%%%%%%%%%%%%%%%%%%%%%%%%%%%%%%%%
\subsection{The Specific Case of BdHNe}
%%%%%%%%%%%%%%%%%%%%%%%%%%%%%%%%%%%%%%%%%%%%%%%%%%%%%%%%%%%%

\begin{figure}[H]
    \centering
    \includegraphics[width=0.7\hsize,clip]{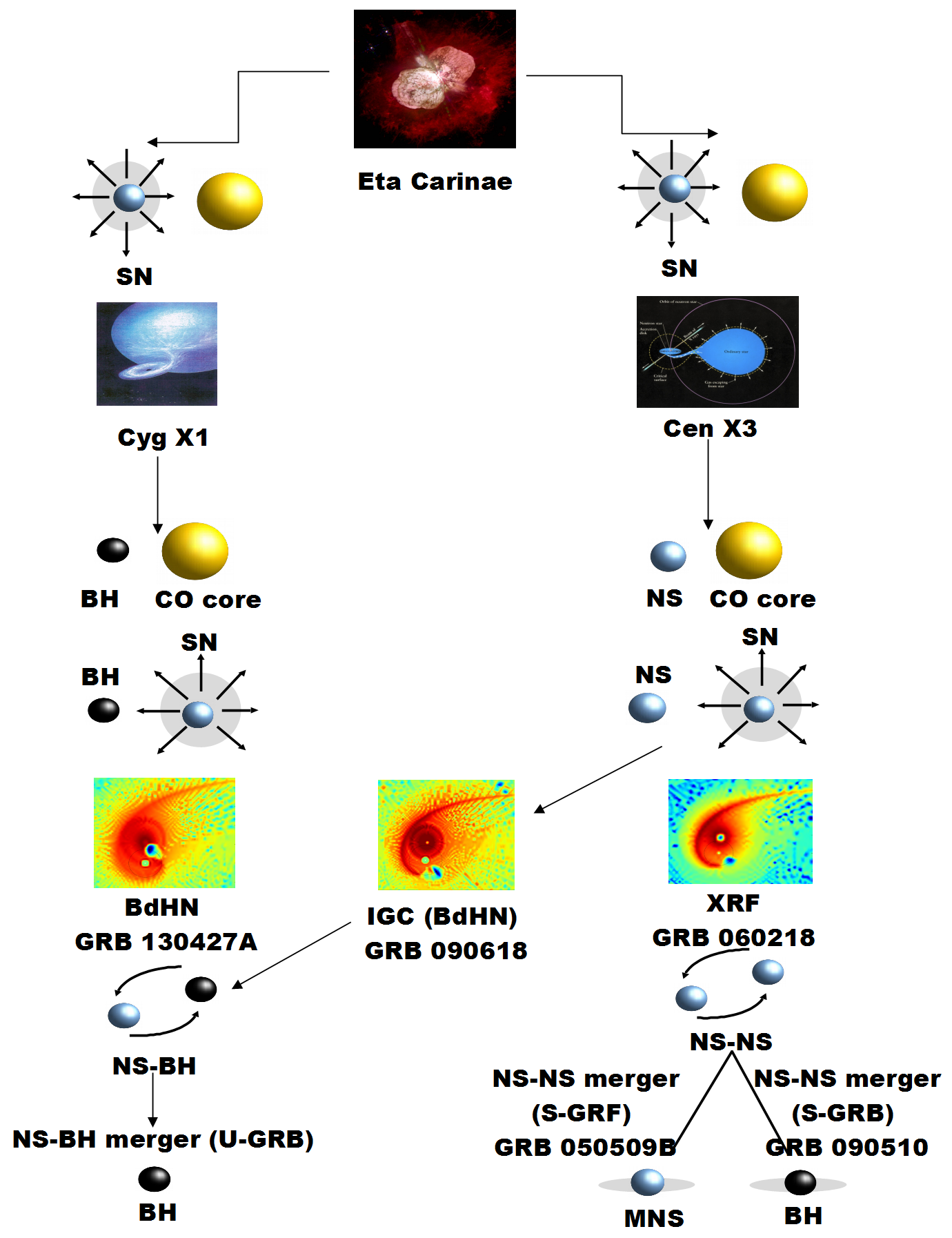}
    \caption{{Taken from Figure~1 in~\cite{2018IJMPA..3344031R}. Binary evolutionary paths leading to BdHNe I (previously named BdHNe) and II (previously named XRFs) and whose out-states, in due time, evolve into progenitors of short GRBs. The massive binary has to survive two core-collapse SN events. The first event forms a~NS (right-side path) or BH (left-side path). The massive companion continues its evolution until it forms a~CO$_{\rm core}$. This simplified evolution diagram which does not show intermediate stages such as common-envelope phases (see e.g.,~\cite{2015PhRvL.115w1102F,2015ApJ...812..100B}, and references therein). At this stage the binary is a~CO$_{\rm core}$-NS (right-side path) or a~CO$_{\rm core}$-BH (left-side path). Then, it occurs the second SN event which forms what we call the $\nu$NS at its center. We focus in this article to review the theoretical and observational aspects of interaction of this SN event with the NS companion (BdHNe I and II). We do not treat here the case of a~SN exploding in an~already formed BH companion (BdHNe IV). At this point the system can form a~$\nu$NS-BH/NS (BdHN I/II) binary (right-side path), or a~$\nu$NS-BH (BdHN IV) in the (left-side path). The emission of gravitational waves will make this compact-object binaries to merge, becoming progenitors of short GRBs~\cite{2015PhRvL.115w1102F}}.
    We recall to the reader that S-GRBs and S-GRFs stand for, respectively, authentic short GRBs and short Gamma-ray flashes, the two subclasses of short bursts from NS-NS mergers, the former produced when the merger leads to a~more massive NS and the latter when a~BH is formed~\cite{2016ApJ...832..136R}.}
    \label{fig:binaryevolution}
\end{figure}

We review in this article the specific case of BdHNe type I and II. As we have mentioned, the progenitor system is an~exploding CO$_{\rm core}$ as a~type Ic SN in presence of a~NS companion~\cite{2016ApJ...832..136R,2019ApJ...874...39W}. Figure~\ref{fig:binaryevolution} shows a~comprehensive summary of the binary path leading to this variety of compact binaries that are progenitors of the above subclasses of long GRBs and that, at the same time, have an~intimate connection with the short GRBs.

\begin{figure}[H]
\centering
\includegraphics[width=0.8\hsize,height=7cm]{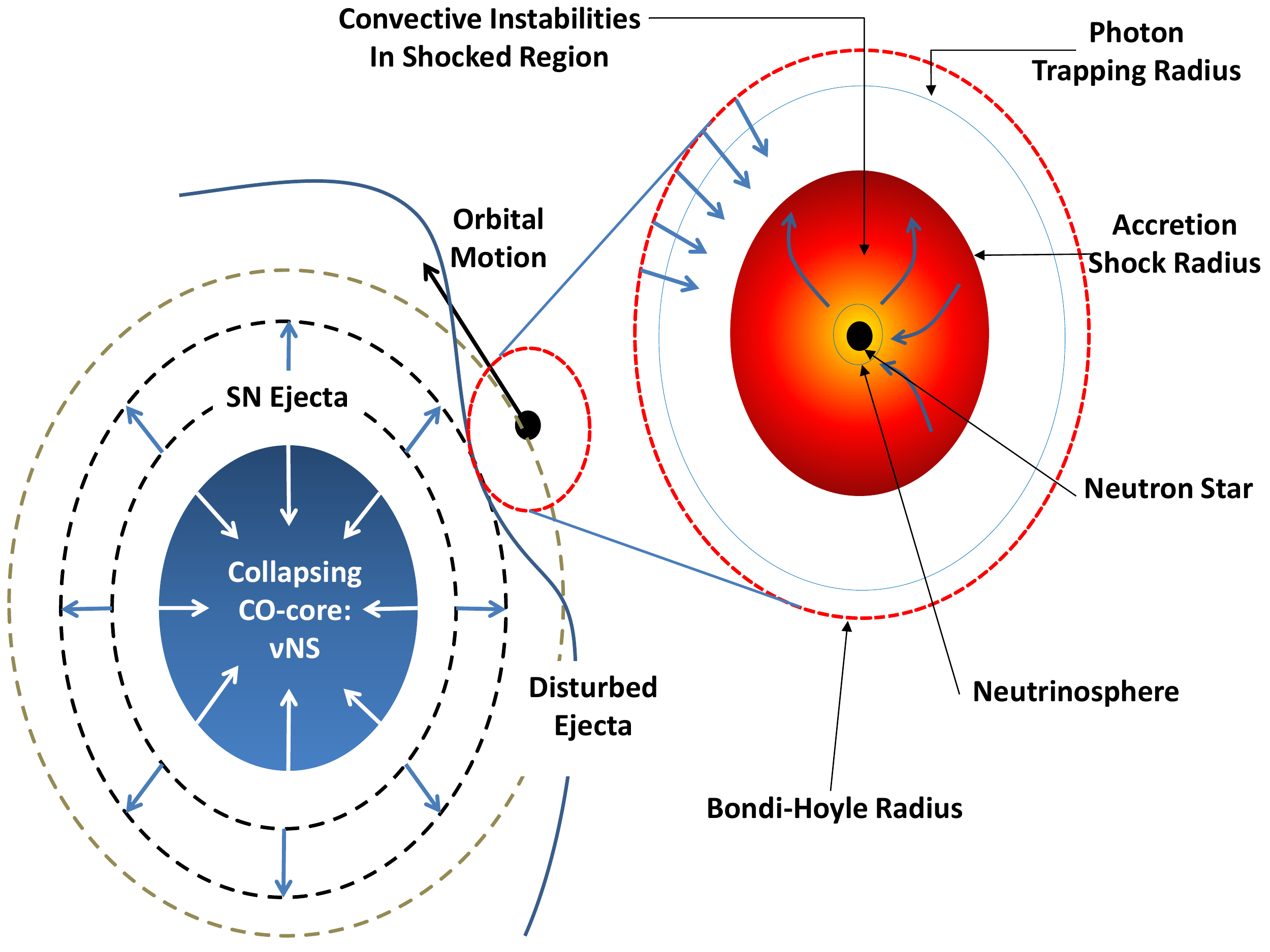}
\caption{Scheme of the induced gravitational collapse (IGC) scenario {(taken from Figure~1 in~\cite{2014ApJ...793L..36F})}. The~CO$_{\rm core}$ undergoes supernova (SN) explosion, the neutron star (NS) accretes part of the SN ejecta and then reaches the critical mass for gravitational collapse to a~black hole (BH), with consequent emission of a~GRB. The SN ejecta reach the NS Bondi-Hoyle radius and fall toward the NS surface. The material shocks and decelerates while it piles over the NS surface. At the neutrino emission zone, neutrinos take away most of the gravitational energy gained by the matter infall. The neutrinos are emitted above the NS surface that allow the material to reduce its entropy to be finally incorporated to the NS. For further details and numerical simulations of the above process see~\cite{2014ApJ...793L..36F,2015ApJ...812..100B,2016ApJ...833..107B}.
}\label{fig:AccretionEsqueme}
\end{figure}

We emphasize on the theoretical framework concerning the CO$_{\rm core}$-NS binaries which have been extensively studied by our group in a~series of publications~\cite{2012ApJ...758L...7R,2012A&A...548L...5I,2014ApJ...793L..36F,2015PhRvL.115w1102F,2015ApJ...812..100B,2016ApJ...833..107B}. The CO$_{\rm core}$ explodes as SN producing an~accretion process onto the NS. For sufficiently compact binaries, e.g., orbital periods of the order of few minutes, the accretion is highly super-Eddington (hypercritical) leading to the possibility of the IGC of the NS once it reaches the critical mass, and forms a~BH (see~Figure~\ref{fig:AccretionEsqueme}).

If the binary is not disrupted by the explosion, BdHNe produces new binaries composed of a~new NS ($\nu$NS) formed at the center of the SN, and a~more massive NS or a~BH companion (see~Figure~\ref{fig:binaryevolution}).

In the case of BH formation, {the rotation of the BH together with the presence of the magnetic field inherited from the NS and the surrounding matter conform to what we have called the \emph{inner engine} of the high-energy emission~\cite{2018arXiv181101839R,2018arXiv181200354R,2019arXiv190404162R,2019arXiv190403163R}. The electromagnetic field of the engine is mathematically described by the Wald solution~\cite{1974PhRvD..10.1680W}. The above ingredients induce an~electric field around the BH which under the BdHN conditions is initially overcritical, creating electron-positron ($e^+e^-$) pair plasma which self-accelerates to ultrarelativistic velocities and whose transparency explains to the GRB prompt emission in Gamma-rays. The electric field is also able to accelerate protons which along the rotation axis lead to ultra high-energy cosmic rays (UHECRs) of up to $10^{21}$~eV. In the other directions the acceleration process lead to proton-synchrotron radiation which explains the GeV emission~\cite{2018arXiv181101839R,2018arXiv181200354R}.} The interaction/feedback of the GRB into the SN makes it become the hypernova (HN)~\cite{2018ApJ...869..151R,2019ApJ...871...14B} observed in the optical, powered by nickel decay, a~few days after the GRB trigger. {The SN shock breakout and the} hypercritical accretion can be observed as X-ray precursors~\cite{2016ApJ...833..107B}. The $e^+e^-$ feedback onto the SN ejecta also produces gamma- and X-ray flares observed in the early afterglow~\cite{2018ApJ...852...53R}. The synchrotron emission by relativistic electrons from the $\nu$NS in the expanding magnetized HN ejecta and the $\nu$NS pulsar emission explain the early and late X-ray afterglow~\cite{2018ApJ...869..101R}.

The article is organized as follows. In Section~\ref{sec:2} we summarized following a~chronological order the (1D, 2D and 3D) numerical simulations of BdHNe up to the year 2016, mentioning their salient features. A detailed explanation of the main ingredients of the calculations (equations of motion, accretion modeling, NS evolution equations, critical mass, accretion-zone hydrodynamics, neutrino emission and accretion energy release) can be found in Section~\ref{sec:3}. The most recent 3D smoothed-particle-hydrodynamics (SPH) numerical simulations of 2018 are presented in Section~\ref{sec:4}. Section~\ref{sec:5} is devoted to the consequences on these simulations on the analysis and interpretation of the GRB multiwavelength data. In Section~\ref{sec:6} we present an~analysis of the binary gravitational binding of BdHNe progenitors, so it is shown that most BdHNe type I are expected to be NS-BH binaries. The cosmological evolutionary scenario leading to the formation of BdHN, their occurrence rate and connection with short GRBs is presented in Section~\ref{sec:7}. %meaning retained?

We show in Table~\ref{tab:acronyms} a~summary of acronyms used in this work.
\begin{table}[H]
\centering
\caption{Acronyms used in this work in alphabetical order.}
\label{tab:acronyms}
\begin{tabular}{lc}
\toprule
\textbf{Extended Wording} & \textbf{Acronym} \\
\midrule
Binary-driven hypernova & BdHN \\
Black hole                    & BH \\
Carbon-oxygen core      & CO$_{\rm core}$ \\ 
Gamma-ray burst         & GRB \\
Gamma-ray flash          & GRF \\
Induced gravitational collapse & IGC \\
Massive neutron star     & MNS \\
Neutron star                & NS \\
New neutron star created in the SN explosion          & $\nu$NS \\
Short Gamma-ray burst  & S-GRB \\
Short Gamma-ray flash  & S-GRF \\
Supernova                  & SN \\
Ultrashort Gamma-ray burst & U-GRB \\
Ultra high-energy cosmic ray & UHECR\\
White dwarf                & WD \\
X-ray flash                  & XRF \\
\bottomrule
\end{tabular}
\end{table}

%%%%%%%%%%%%%%%%%%%%%%%%%%%%%%%%%%%%%%%%%%%%%%%%%%%%%%%%%%%%
%%%%%%%%%%%%%%%%%%%%%%%%%%%%%%%%%%%%%%%%%%%%%%%%%%%%%%%%%%%%
\section{A Chronological Summary of the IGC Simulations: 2012--2016}\label{sec:2}
%%%%%%%%%%%%%%%%%%%%%%%%%%%%%%%%%%%%%%%%%%%%%%%%%%%%%%%%%%%%
%%%%%%%%%%%%%%%%%%%%%%%%%%%%%%%%%%%%%%%%%%%%%%%%%%%%%%%%%%%%

%%%%%%%%%%%%%%%%%%%%%%%%%%%%%%%%%%%%%%%%%%%%%%%%%%%%%%%%%%%%
\subsection{First Analytic Estimates}
%%%%%%%%%%%%%%%%%%%%%%%%%%%%%%%%%%%%%%%%%%%%%%%%%%%%%%%%%%%%

The IGC scenario was formulated in 2012~\cite{2012ApJ...758L...7R} presenting a~comprehensive astrophysical picture supporting this idea as well as a~possible evolutionary scenario leading to the progenitor CO$_{\rm core}$-NS binaries. It was also there presented an~analytic formula for the accretion rate onto the NS companion on the basis of the following simplified assumptions: (1) a~uniform density profile of the pre-SN CO$_{\rm core}$; (2) the ejecta was evolved following an~homologous expansion; (3) the mass of the NS (assumed to be initially $1.4~M_\odot$) and the CO$_{\rm core}$ (in the range $4$--$8~M_\odot$) were assumed nearly constant. So, it was shown that the accretion rate onto the NS is highly super-Eddington, namely it is hypercritical, reaching values of up to $0.1~M_\odot$~s$^{-1}$ for compact binaries with orbital periods of the order of a~few minutes. This estimate implied that the hypercritical accretion could induce the gravitational collapse of the NS which, in a~few seconds, would reach the critical mass with consequence formation of a~BH. A first test of this IGC first model in real data was soon presented in the case of GRB 090618~\cite{2012A&A...548L...5I}.

\subsection{First Numerical Simulations: 1D Approximation}

\begin{figure}[H]
\centering
\includegraphics[width=0.6\hsize,clip]{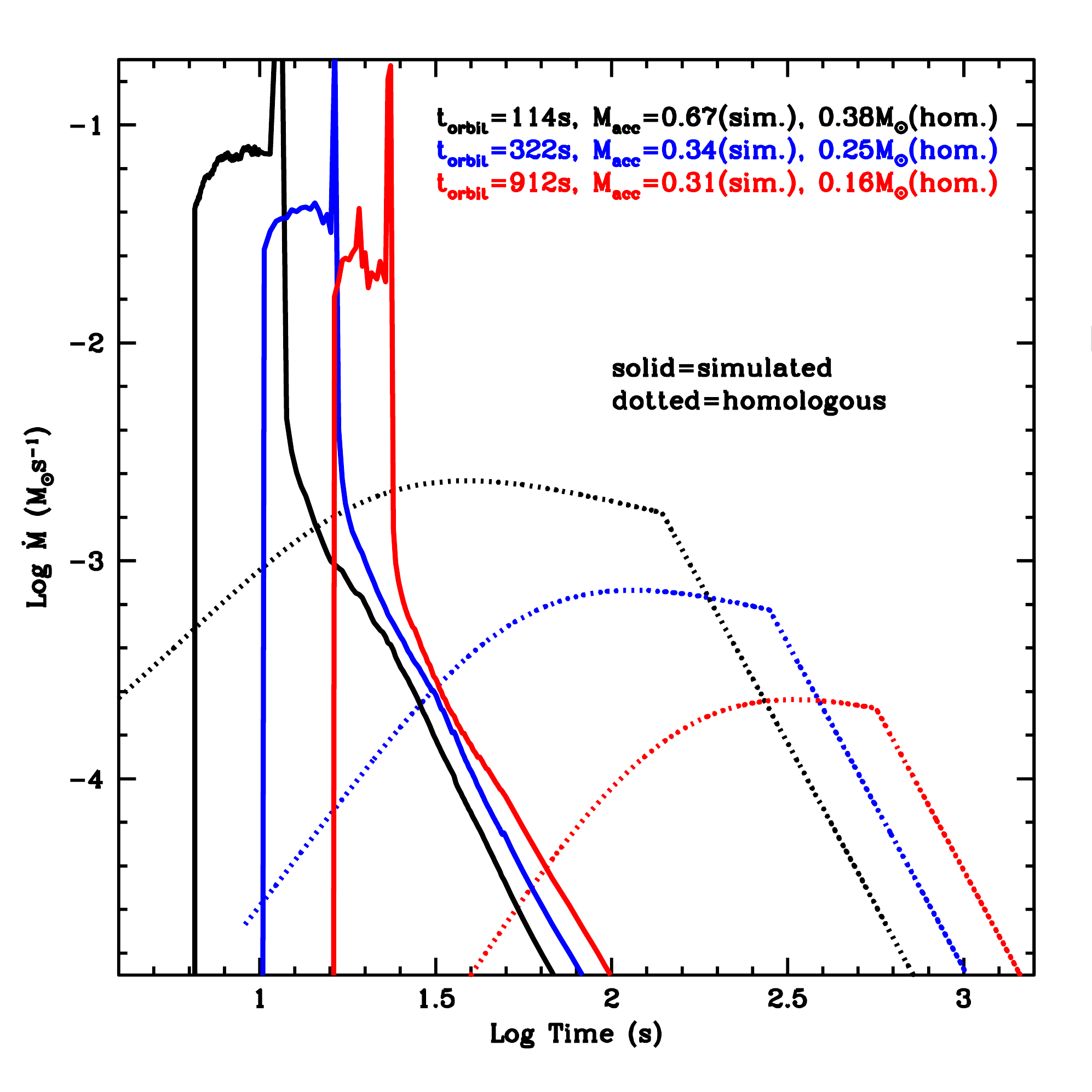}
\caption{Hypercritical accretion {rate onto the NS companion} for selected separation distances. The~CO$_{\rm core}$ is obtained with a~progenitor star of zero-age main-sequence (ZAMS) mass of $20\,M_\odot$, calculated in~\cite{2014ApJ...793L..36F}. The numerical calculation leads to a~sharper accretion profile with respect to the one obtained assuming homologous expansion of the SN ejecta. {Taken from Figure~3 in~\cite{2014ApJ...793L..36F}}.}\label{fig:fryer2014}
\end{figure}

The first numerical simulations were implemented in 2014 in~\cite{2014ApJ...793L..36F} via a~1D code including (see~Figure~\ref{fig:fryer2014}): (1) the modeling of the SN via the 1D core-collapse SN code of Los Alamos~\cite{1999ApJ...516..892F}; (2) the microphysics experienced by the inflow within the accretion region including the neutrino ($\nu$) emission and hydrodynamics processes such as shock formation; (3) with the above it was followed by the evolution of the material reaching the Bondi-Hoyle capture region and the subsequent in-fall up to the NS surface. Hypercritical accretion rates in the range $10^{-3}$--$10^{-1}~M_\odot$~s$^{-1}$ were inferred, confirming the first analytic estimates and the IGC of the NS companion for binary component masses similar to the previous ones and for orbital periods of the order of $5$~min. %meaning retained?

The above simulations were relevant in determining that the fate of the system is mainly determined by the binary period ($P$); the SN ejecta velocity ($v_{\rm ej}$) and the NS initial mass.~$P$ and $v_{\rm ej}$ enter explicitly in the Bondi-Hoyle accretion rate formula through the capture radius expression, and implicitly via the ejecta density since they influence the decompression state of the SN material at the NS position.

\subsection{2D Simulations including Angular Momentum Transfer}

\begin{figure}[H]
\centering
\includegraphics[scale=0.32]{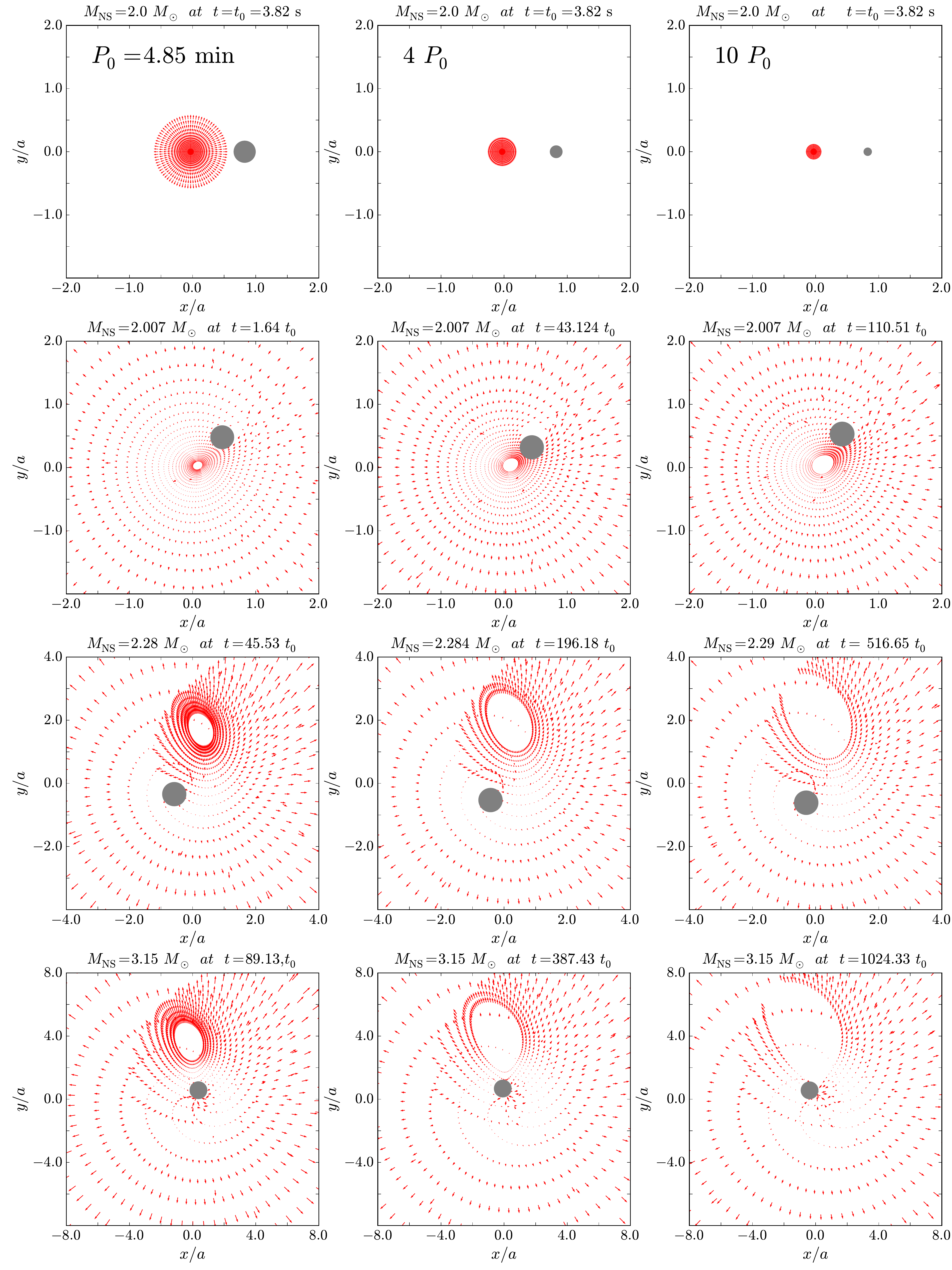}
\caption{{Numerical simulations of the SN ejecta velocity field (red arrows)} at selected times of the accretion process onto the NS {(taken from Figure~3 in~\cite{2015ApJ...812..100B}}). In these snapshots we have adopted the CO$_\textrm{core}$ obtained from a~$M_{\rm ZAMS}=30~M_\odot$ progenitor; an~ejecta outermost layer velocity \mbox{$v_{0_{\rm star}}=2\times 10^9$~cm~s$^{-1}$}, an~initial NS mass, $M_{\rm NS}(t=t_0)=2.0~M_\odot$.~The minimum orbital period to have no Roche-lobe overflow is $P_0=4.85$~min. In the left, central and right columns of snapshots we show the results for binary periods $P=P_0$, $4 P_0$, and $10 P_0$, respectively. The Bondi-Hoyle surface, the filled gray circle, increases as the evolution continues mainly due to the increase of the NS mass {(the decrease of the lower panels is only apparent due to the enlargement of the x-y scales)}. The x-y positions refer to the center-of-mass reference frame. The last image in each column corresponds to the instant when the NS reaches the critical mass value. For the initial conditions of these simulations, the~NS ends its evolution at the mass-shedding limit with a~maximum value of the angular momentum $J=6.14\times 10^{49}$~g~cm$^2$~s$^{-1}$ and a~corresponding critical mass of $3.15~M_\odot$.}\label{fig:vfield}
\end{figure}

Soon after, in 2015, we implemented in~\cite{2015ApJ...812..100B} a~series of improvements to the above calculations by relaxing some of the aforementioned assumptions (see~Figure~\ref{fig:vfield}). We adopted for the ejecta a~density profile following a~power-law with the radial distance and evolved it with an~homologous expansion. The angular momentum transport, not included in the previous estimates, was included. With this addition it was possible to estimate the spin-up of the NS companion by the transfer of angular momentum from the in-falling matter which was shown to circularize around the NS before being accreted. General relativistic effects were also introduced, when calculating the evolution of the structure parameters (mass, radius, spin, etc) of the accreting NS, in the NS gravitational binding energy, and in the angular momentum transfer by the circularized particles being accreted from the innermost circular orbit.

One of the most important results of~\cite{2015ApJ...812..100B} was that, taking into account that the longer the orbital $P$ the lower the accretion rate, it was there computed the maximum orbital period ($P_{\rm max}$) for which the NS reaches the critical mass for gravitational collapse, so for BH formation. The dependence of $P_{\rm max}$ on the initial mass of the NS was also there explored. The orbital period $P_{\rm max}$ was then presented as the separatrix of two families of long GRBs associated with these binaries: at the time we called them \emph{Family-1}, the systems in which the NS does not reach the critical mass, and \emph{Family-2} the ones in which it reaches the critical mass and forms a~BH. It can be seen that the Family-1 and Family-2 long GRBs evolve subsequently into the concepts of \emph{XRFs} and \emph{BdHNe}, respectively.

\subsection{First 3D Simulations}

A great step toward the most recent simulations was achieved in 2016 in~\cite{2016ApJ...833..107B} where an~SPH-like simulation was implemented in which the SN ejecta was emulated by ``point-like'' particles. The mass and number of the particles populating each layer were assigned, for self-consistency, according to the power-law density profile. The initial velocity of the particles of each layer was set, in agreement with the chosen power-law density profile, following a~radial velocity distribution; i.e.,~$v\propto r$. 

The evolution of the SN particles was followed by Newtonian equations of motion in the gravitational field of the NS companion, also taking into account the orbital motion which was included under the assumption that the NS performs a~circular orbit around the CO$_{\rm core}$ center that acts as the common center-of-mass, namely assuming that the mass of the pre-SN core is much larger than the NS mass.

The accretion rate onto the NS was computed, as in~\cite{2015ApJ...812..100B}, using the Bondi-Hoyle accretion formula and, every particle reaching the Bondi-Hoyle surface, was removed from the system. The maximum orbital period $P_\textrm{max}$ in which the NS collapses by accretion could be further explored including the dependence on the mass of the pre-SN CO$_\textrm{core}$, in addition to the dependence on the NS mass.

A detailed study of the hydrodynamics and the neutrino emission in the accretion region on top the NS surface was performed. Concerning the neutrino emission, several $\nu$ and antineutrino ($\bar{\nu}$) production processes were considered and showed that electron-positron annihilation ($e^+e^- \to \nu\bar{\nu}$) overcomes by orders of magnitude any other mechanism of neutrino emission in the range of accretion rates $10^{-8}$--$10^{-2}~M_\odot$~s$^{-1}$, relevant for XRFs and BdHNe. The neutrino luminosity can reach values of up to $10^{52}$~erg~s$^{-1}$ and the neutrino mean energy of $20$~MeV for the above upper value of the accretion rate. For the reader interested in the neutrino emission, we refer to~\cite{2018ApJ...852..120B} for a~detailed analysis of the neutrino production in XRFs and BdHNe including flavor oscillations experienced by the neutrinos before abandoning the system.

Concerning the hydrodynamics, the evolution of the temperature and density of outflows occurring during the accretion process owing to convective instabilities was estimated. It was there shown the interesting result that the temperature of this outflow and its evolution can explain the early (i.e.,~precursors) X-ray emission that has been observed in some BdHNe and in XRFs, exemplified there analyzing the early X-ray emission observed in GRB 090618, a~BdHN I, and in GRB 060218, a~BdHN II (an XRF).

\begin{figure}[H]
    \centering
    \includegraphics[width=0.8\hsize,clip]{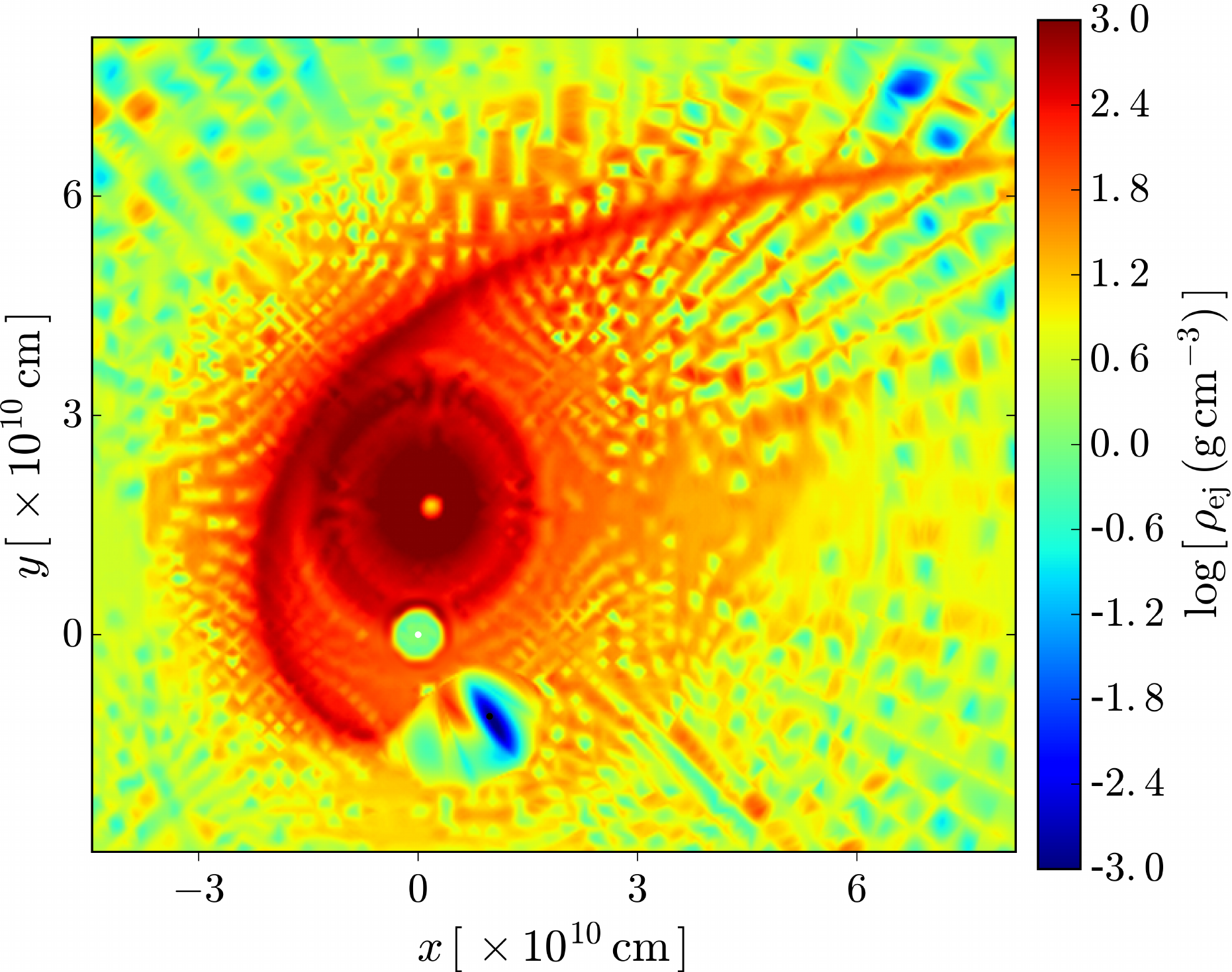}%    \includegraphics[width=0.48\hsize,clip]{mass_profiles.pdf}
    \caption{Snapshot of the SN ejecta density in the orbital plane of the CO$_{\rm core}$-NS binary. {Numerical simulation taken from Figure~6 in~\cite{2016ApJ...833..107B}}. The plot corresponds to the instant when the NS reaches the critical mass and forms the BH (black dot), approximately $250$~s from the SN explosion. The $\nu$NS is represented by the white dot. The binary parameters are: the initial mass of the NS companion is $2.0~M_\odot$; the CO$_{\rm core}$ leading to an~ejecta mass of $7.94~M_\odot$, and the orbital period is $P\approx 5$~min, namely a~binary separation $a\approx 1.5\times 10^{10}$~cm.}
    \label{fig:profiles}
\end{figure}

A most important result of these simulations was the possibility of having a~first glance of the morphology acquired by the SN ejecta: the matter density, initially spherically symmetric, becomes highly asymmetric due to the accretion process and the action of the gravitational field of the NS companion (see~Figure~\ref{fig:profiles}). 

%%%%%%%%%%%%%%%%%%%%%%%%%%%%%%%%%%%%%%%%%%%%%%%%%%%%%%%%%%%%%%%%
%%%%%%%%%%%%%%%%%%%%%%%%%%%%%%%%%%%%%%%%%%%%%%%%%%%%%%%%%%%%%%%%
\section{The Hypercritical Accretion Process}\label{sec:3}
%%%%%%%%%%%%%%%%%%%%%%%%%%%%%%%%%%%%%%%%%%%%%%%%%%%%%%%%%%%%%%%%
%%%%%%%%%%%%%%%%%%%%%%%%%%%%%%%%%%%%%%%%%%%%%%%%%%%%%%%%%%%%%%%%

We now give details of the accretion process within the IGC scenario following~\cite{2014ApJ...793L..36F,2015ApJ...812..100B,2015PhRvL.115w1102F,2016ApJ...833..107B}. There are two main physical conditions for which hypercritical (i.e., highly super-Eddington) accretion onto the NS occurs in XRFs and BdHNe. The first is that the photons are trapped within the inflowing material and the second is that the shocked atmosphere on top of the NS becomes sufficiently hot ($T\sim 10^{10}$~K) and dense ($\rho \gtrsim 10^6$~g~cm$^{-3}$) to produce a~very efficient neutrino-antineutrino ($\nu\bar\nu$) cooling emission. In this way the neutrinos become mainly responsible for releasing the energy gained by accretion, allowing hypercritical accretion to continue.

\subsection{Accretion Rate and NS Evolution}

The first numerical simulations of the IGC were performed in~\cite{2014ApJ...793L..36F}, including: (1) realistic SN explosions of the CO$_{\rm core}$; (2) the hydrodynamics within the accretion region; (3) the simulated evolution of the SN ejecta up to their accretion onto the NS.
\mbox{Becerra et al.~\cite{2015ApJ...812..100B}} then estimated the amount of angular momentum carried by the SN ejecta and how much is transferred to the NS companion by accretion. They showed that the SN ejecta can circularize for a~short time and form a~disc-like structure surrounding the NS before being accreted. The evolution of the NS central density and rotation angular velocity (the NS is spun up by accretion) was computed from full numerical solutions of the axisymmetric Einstein equations. The unstable limits of the NS are set by the mass-shedding (or Keplerian) limit and the critical point of gravitational collapse given by the secular axisymmetric instability, (see, e.g.,~\cite{2015ApJ...812..100B} for~details).

The accretion rate of the SN ejecta onto the NS is given by:
\begin{equation}\label{eq:BondiMassRate_definition}
\begin{array}{rclcrcl}
\dot{M}_B(t)&=&\pi\rho_{\rm ej}R_{{\rm cap}}^2\sqrt{v_{{\rm rel}}^2+c_{\rm s,ej}^2}, \qquad R_{{\rm cap}}(t)&=&\cfrac{2 G M_{{\rm NS}}(t)}{v_{{\rm rel}}^2+c_{{\rm s,ej}}^2},
 \end{array}
\end{equation}
where $G$ is the gravitational constant, $\rho_{\rm ej}$ and $c_{{\rm s,ej}}$ are the density and sound speed of the ejecta, $R_{\rm cap}$ and $M_{\rm NS}$ are the NS gravitational capture radius (Bondi-Hoyle radius) and gravitational mass, and~$v_{{\rm rel}}$ the ejecta velocity relative to the NS: $\vec{v}_{{\rm rel}}=\vec{v}_{{\rm orb}}-\vec{v}_{{\rm ej}}$; $|\vec{v}_{\rm orb}|=\sqrt{G(M_{\rm core}+M_{\rm NS})/a}$, and $\vec{v}_{\rm ej}$ is the velocity of the supernova ejecta (see~Figure~\ref{fig:AccretionEsqueme}).

Numerical simulations of the SN explosions suggest the adopted homologous expansion of the SN, i.e.,~$v_{\rm ej}(r,t)=n r/t$, where $r$ is the position of each layer from the SN center and $n$ is the expansion parameter. The density evolves as 
\begin{equation}
\rho_{\rm ej}(r,t)=\rho_{\rm ej}^0(r/R_{\rm star}(t),t_0)\frac{M_{\rm env}(t)}{M_{\rm env}(0)}\left(\frac{R_{\rm star}(0)}{R_{\rm star}(t)}\right)^3,
\end{equation}
where $M_{\rm env}(t)$ the mass of the CO$_{\rm core}$ envelope, $R_{\rm star}(t)$ is the radius of the outermost layer, and $\rho_{\rm ej}^0$ is the pre-SN CO$_{\rm core}$ density profile; $\rho_{\rm ej}(r,t_0)= \rho_{\rm core} (R_{\rm core}/r)^m$, where $\rho_{\rm core}$, $R_{\rm core}$ and $m$ are the profile parameters obtained from numerical simulations. Typical parameters of the CO$_{\rm core}$ mass are (3.5--9.5)~$M_\odot$ corresponding to $(15$--$30)~M_\odot$ zero-age-main-sequence (ZAMS) progenitors (see~\cite{2014ApJ...793L..36F,2015ApJ...812..100B} for details). The binary period is limited from below by the request of having no Roche lobe overflow by the CO$_{\rm core}$ before the SN explosion~\cite{2014ApJ...793L..36F}. For instance, for a~CO$_{\rm core}$ of 9.5~$M_\odot$ forming a~binary system with a~2~$M_\odot$ NS, the minimum orbital period allowed by this condition is $P_{\rm  min}\approx 5$~min. For~these typical binary and pre-SN parameters, Equation~(\ref{eq:BondiMassRate_definition}) gives accretion rates $10^{-4}$--$10^{-2} M_\odot$~s$^{-1}$.

We adopt an~initially non-rotating NS companion so its exterior spacetime at time $t=0$ is described by the Schwarzschild metric. The SN ejecta approach the NS with specific angular momentum, $l_{\rm acc}=\dot{L}_{\rm cap}/\dot{M}_B$, circularizing at a~radius $r_{\rm circ}\geq r_{\rm lco}$  if $l_{\rm acc} \geq l_{\rm lso}$ with $r_{\rm lco}$ the radius of the last circular orbit (LCO). For a~non-rotating NS $r_{\rm lco}=6 G M_{\rm NS}/c^2$ and $l_{\rm lco}=2\sqrt{3} G M_{\rm NS}/c$. For typical parameters, $r_{\rm circ}/r_{\rm lco}\sim 10$--$10^3$. 

The accretion onto the NS proceeds from the radius $r_{\rm in}$. The NS mass and angular angular momentum evolve as~\cite{2015ApJ...812..100B,2017PhRvD..96b4046C}:
\begin{equation}\label{eq:AngMom}
\dot{M}_{\rm NS}=\left(\frac{\partial M_{\rm NS}}{\partial M_b}\right)_{J_{\rm NS}}\dot{M}_b+\left(\frac{\partial M_{\rm NS}}{\partial J_{\rm NS}}\right)_{M_b}\dot{J}_{\rm NS},\qquad \dot{J}_{\rm NS}=\xi \, l(r_{\rm in})\dot{M}_{\rm B},
\end{equation}
where $M_b$ is the NS baryonic mass, $l(r_{\rm in})$ is the specific angular momentum of the accreted material at $r_{\rm in}$, which corresponds to the angular momentum of the LCO, and $\xi\leq 1$ is a~parameter that measures the efficiency of angular momentum transfer. In this picture we have $\dot{M}_b = \dot{M}_B$.

For the integration of Equations~(\ref{eq:BondiMassRate_definition}) and (\ref{eq:AngMom}) we have to supply the values of the two partial derivatives in Equation~(\ref{eq:AngMom}). They are obtained from the relation of the NS gravitational mass, $M_{\rm NS}$, with $M_b$ and $J_{\rm NS}$, namely from the knowledge of the NS binding energy. For this we use the general relativistic calculations of rotating NSs presented in~\cite{2015PhRvD..92b3007C}. They show that, independent on the nuclear EOS, the following analytical formula represents the numerical results with sufficient accuracy \mbox{(error $<2\%$)}:
\begin{equation}\label{eq:MbMnsjns}
\frac{M_b}{M_\odot}=\frac{M_{\rm NS}}{M_\odot}+\frac{13}{200}\left(\frac{M_{\rm NS}}{M_\odot}\right)^2\left(1-\frac{1}{137}j_{\rm NS}^{1.7}\right),
\end{equation}
where $j_{\rm NS}\equiv cJ_{\rm NS}/(G M_\odot^2)$. 

In the accretion process, the NS gains angular momentum and therefore spins up. To evaluate the amount of angular momentum transferred to the NS at any time we include the dependence of the LCO specific angular momentum as a~function of $M_{\rm NS}$ and $J_{\rm NS}$. For corotating orbits, the following relation is valid for the NL3, TM1 and GM1 EOS~\cite{2017PhRvD..96b4046C,2015ApJ...812..100B}:
\begin{equation}
l_{\rm lco}= \frac{G M_{\rm NS}}{c}\left[2 \sqrt{3} - 0.37 \left(\frac{j_{\rm NS}}{M_{\rm NS}/M_\odot}\right)^{0.85}\right].
\end{equation}

The NS continues the accretion until it reaches an~instability limit or up to when all the SN ejecta overcomes the NS Bondi-Hoyle region. We take into account the two main instability limits for rotating NSs: the mass-shedding or Keplerian limit and the secular axisymmetric instability limit. The latter defines critical NS mass. For the aforementioned nuclear EOS, the critical mass can be approximately written as~\cite{2015PhRvD..92b3007C}:
\begin{equation}\label{eq:Mcrit}
M_{\rm NS}^{\rm crit}=M_{\rm NS}^{J=0}(1 + k j_{\rm NS}^p),
\end{equation}
where $k$ and $p$ are EOS-dependent parameters (see Table~\ref{tb:StaticRotatingNS}). These formulas fit the numerical results with a~maximum error of 0.45\%.
\begin{table}[H]
\centering
\caption{Critical NS mass in the non-rotating case and constants  $k$ and $p$ needed to compute the NS critical mass in the non-rotating case given by Equation~(\ref{eq:Mcrit}). The values are for the NL3, GM1 and TM1~EOS.}
{\begin{tabular}{@{}cccc@{}} 
\toprule
\textbf{EOS}  &  \boldmath$M_{\rm crit}^{J=0}$~$(M_{\odot})$ & \boldmath$p$&\boldmath{$k$} \\\midrule
NL3 & $2.81$ & $1.68$ & $0.006$\\
GM1 & $2.39$ & $1.69$ & $0.011$\\
TM1 & $2.20$ & $1.61$ & $0.017$\\
\bottomrule
\end{tabular}}
\label{tb:StaticRotatingNS}
\end{table}

Additional details and improvements of the hypercritical accretion process leading to XRFs and BdHNe were presented in~\cite{2016ApJ...833..107B}. Specifically:
\begin{enumerate}
\item
The density profile included finite size/thickness effects and additional CO$_{\rm core}$ progenitors, leading to different SN ejecta masses being considered.
\item
In~\cite{2015ApJ...812..100B} the maximum orbital period, $P_{\rm max}$, over which the accretion onto NS companion is not sufficient to bring it to the critical mass, was inferred. Thus, binaries with $P > P_{\rm max}$ lead to XRFs while the ones with $P\lesssim P_{\rm max}$ lead to BdHNe. \mbox{Becerra et al.~\cite{2016ApJ...833..107B}} extended the determination of $P_{\rm max}$ for all the possible initial values of the NS mass. They also examined the outcomes for different values of the angular momentum transfer efficiency parameter.
\item
The expected luminosity during the process of hypercritical accretion for a~wide range of binary periods covering both XRFs and BdHNe was estimated.
\item
It was shown that the presence of the NS companion originates asymmetries in the SN ejecta (see,~e.g.,~Figure~6 in~\cite{2016ApJ...833..107B}). The signatures of such asymmetries in the X-ray emission was there shown in the specific example of XRF 060218.
\end{enumerate}

%%%%%%%%%%%%%%%%%%%%%%%%%%%%%%%%%%%%%%%%%%%%%%%%%%%%%%%%%%%%%%%%
\subsection{Hydrodynamics in the Accretion Region}
%%%%%%%%%%%%%%%%%%%%%%%%%%%%%%%%%%%%%%%%%%%%%%%%%%%%%%%%%%%%%%%%

The accretion rate onto the NS can be as high as $\sim$10$^{-2}$--$10^{-1}~M_{\odot}$~s$^{-1}$. For such accretion rates:
\begin{enumerate}
\item 
The magnetic pressure is much smaller than the random pressure of the infalling material, therefore the magnetic-field effects on the accretion process are negligible~\cite{1996ApJ...460..801F,2012ApJ...758L...7R}.
\item 
The photons are trapped within the infalling matter, hence the Eddington limit does not apply and hypercritical accretion occurs. The trapping radius is defined as~\cite{1989ApJ...346..847C}: \mbox{$r_{\rm trapping}={\rm min}\{\dot{M}_B\kappa/(4\pi c),R_{\rm cap}\}$}, where $\kappa$ is the opacity.~\cite{2014ApJ...793L..36F} estimated a~Rosseland mean opacity of $\approx$5 $\times$ 10$^3$~cm$^2$~g$^{-1}$ for the CO$_{\rm cores}$. This, together with our typical accretion rates, lead~to $\dot{M}_B\kappa/(4\pi c)\sim 10^{13}$--$10^{19}$~cm. This radius is much bigger than the Bondi-Hoyle radius. 
\item 
The above condition, and the temperature-density values reached on top of the NS surface, lead to an~efficient neutrino cooling which radiates away the gain of gravitational energy of the infalling material~\cite{1972SvA....16..209Z,1973PhRvL..31.1362R,1996ApJ...460..801F,2012ApJ...758L...7R,2014ApJ...793L..36F}.
\end{enumerate}

The accretion shock moves outward as the material piles onto the NS. Since the post-shock entropy is inversely proportional to the shock radius position, the NS atmosphere is unstable with respect to Rayleigh-Taylor convection at the beginning of the accretion process. Such instabilities might drive high-velocity outflows from the accreting NS~\cite{2006ApJ...646L.131F,2009ApJ...699..409F}. The entropy at the base of the atmosphere is~\cite{1996ApJ...460..801F}: 
\begin{equation}
S_{\rm bubble} \approx 16\left(\frac{1.4\,M_\odot}{M_{\rm NS}}\right)^{-7/8}\left(\frac{M_\odot\,{\rm s}^{-1}}{\dot{M}_{\rm B}}\right)^{1/4}\left(\frac{10^6\, {\rm cm}}{r}\right)^{3/8}\,k_B/{\rm nucleon},
\end{equation}
where $k_B$ is the Boltzmann constant. The material expands and cools down adiabatically, i.e.,~$T^3/\rho$ = constant. In the case of a~spherically symmetric expansion, $\rho \propto 1/r^3$ and \mbox{$k_B T_{\rm bubble}=195\, S_{\rm bubble}^{-1}\left(10^6\, {\rm cm}/r\right)$~MeV}. In the more likely case that the material expand laterally we have~\cite{2009ApJ...699..409F}: $\rho \propto 1/r^2$, i.e., $T_{\rm bubble} = T_0 (S_{\rm bubble}) \left(r_0/r\right)^{2/3}$, where $T_0(S_{\rm bubble})$ is obtained from the above equation at $r=r_0\approx R_{\rm NS}$. This implies a~bolometric blackbody flux at the source from the rising~bubbles:
\begin{equation}\label{eq:Lbubble}
F_{\rm bubble} \approx 2\times 10^{40} \left(\frac{M_{\rm NS}}{1.4\,M_\odot} \right)^{-7/2}\left( \frac{\dot{M}_{\rm B}}{M_\odot\,{\rm s}^{-1}} \right)\left( \frac{R_{\rm NS}}{10^6\,{\rm cm}} \right)^{3/2}\left(\frac{r_0}{r}\right)^{8/3}\,{\rm erg\,s}^{-1}{\rm cm}^{-2}.
\end{equation}

\begin{figure}[H]
    \centering
    \includegraphics[width=0.9\hsize,clip]{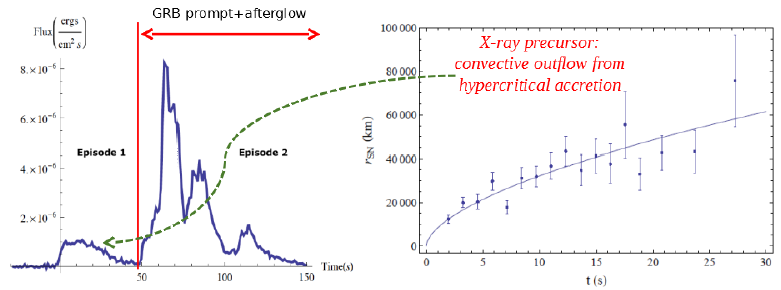}
    \caption{(\textbf{a}) Fermi-GBM (NaI $8$--$440$~keV) light-curve of GRB 090618 {(adapted from Figure~1 in~\cite{2012A&A...543A..10I})}. (\textbf{b}) Expanding radius of the thermal blackbody emission observed in the ``Episode 1'' of GRB 090618 {(adapted from Figure~2 in~\cite{2012A&A...543A..10I})}. The interpretation of such an~X-ray precursor as being due to the emission of the convective bubbles during the process of hypercritical accretion onto the NS was proposed for the first time in~\cite{2014ApJ...793L..36F}.}
    \label{fig:Ep1GRB090618}
\end{figure}

The above thermal emission has been shown~\cite{2014ApJ...793L..36F} to be a~plausible explanation of the early X-ray (precursor) emission observed in some GRBs. The X-ray precursor observed in GRB 090618~\cite{2012A&A...543A..10I,2012A&A...548L...5I} is explained adopting an~accretion rate of $10^{-2}~M_\odot$~s$^{-1}$, the bubble temperature drops from $50$~keV to $15$~keV while expanding from $r\approx 10^9$~cm to $6\times 10^9$~cm (see~Figure~\ref{fig:Ep1GRB090618}). More recently, the X-ray precursor has been observed in GRB 180728A and it is well explained by a~bubble of  $\sim$7~keV at $\sim$$10^{10}$~cm and an~accretion rate of $10^{-3}~M_\odot$~s$^{-1}$ (see~\cite{2019ApJ...874...39W} for details).

%%%%%%%%%%%%%%%%%%%%%%%%%%%%%%%%%%%%%%%%%%%%%%%%%%%%%%%%%%%%%%%%
\subsection{Neutrino Emission and Effective Accretion Rate}
%%%%%%%%%%%%%%%%%%%%%%%%%%%%%%%%%%%%%%%%%%%%%%%%%%%%%%%%%%%%%%%%

\begin{figure}[H]
    \centering
    \includegraphics[width=0.5\hsize,clip]{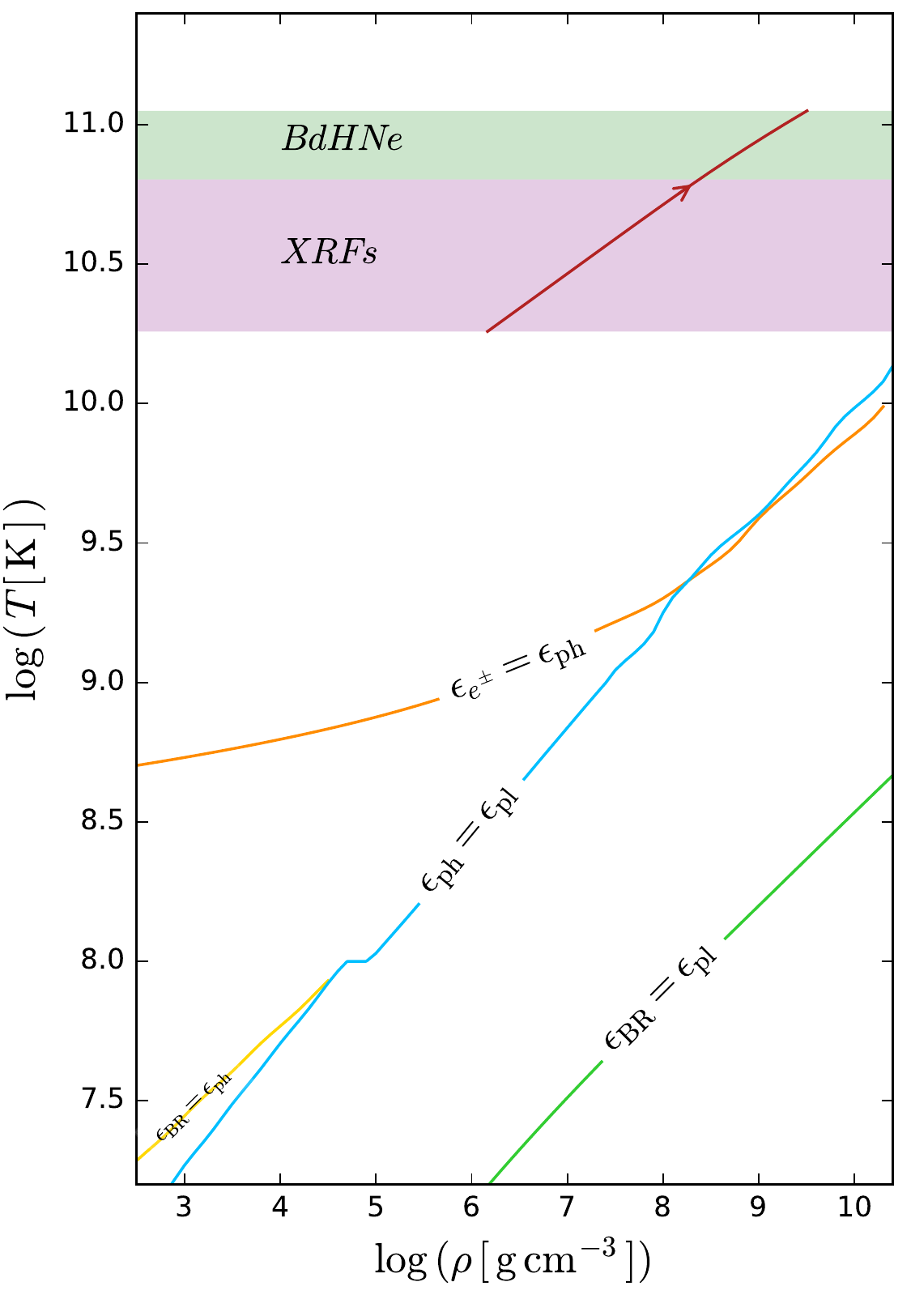}
    \caption{Temperature-density reached by the accreting atmosphere {(taken from Figure~16 in~\cite{2016ApJ...833..107B})}. The~contours indicate where the emissivity of the different neutrinos processes becomes quantitatively equal to each other. We considered the following processes: pair annihilation ($\epsilon_{e^\pm}$), photo-neutrino emission ($\epsilon_{\gamma}$), plasmon decay ($\epsilon_\textrm{pl}$) and Bremsstralung emission ($\epsilon_\textrm{BR}$). The solid red curve spans $T-rho$ values corresponding to accretion rates from $10^{-8}$ to $10^{-1}~M_\odot$~s$^{-1}$ from the lower to the upper end. For any accretion rate of interest, the electron-positron pair annihilation dominates the neutrino emission.}
    \label{fig:Trho}
\end{figure}

For the accretion rate conditions characteristic of our models at peak $\sim$10$^{-4}$--$10^{-2}~M_\odot$~s$^{-1}$, pair~annihilation dominates the neutrino emission and electron neutrinos remove the bulk of the energy~\cite{2016ApJ...833..107B}. The $e^+e^-$ pairs producing the neutrinos are thermalized at the matter temperature. This~temperature is approximately given by:

\begin{equation} \label{eq:Tacc}
T_{\rm acc}\approx \left(\frac{3 P_{\rm shock}}{4 \sigma/c}\right)^{1/4}=\left(\frac{7}{8} \frac{\dot{M}_{\rm acc}
v_{\rm acc} c}{4 \pi R^2_{\rm NS} \sigma}\right)^{1/4},
\end{equation}
where $P_{\rm shock}$ is the pressure of the shock developed on the accretion zone above the NS surface, $\dot{M}_{\rm acc}$~is the accretion rate, $v_{\rm acc}$ is the velocity of the infalling material, $\sigma$ is the Stefan-Boltzmann constant and $c$ the speed of light. It can be checked that, for the accretion rates of interest, the system develops temperatures and densities $T\gtrsim 10^{10}$~K and $\rho\gtrsim 10^6$~g~cm$^{-3}$; respectively (see~Figure~\ref{fig:Trho}).

Under these conditions of density and temperature the neutrino emissivity of the $e^+e^-$ annihilation process can be estimated by the simple formula~\cite{2001PhR...354....1Y}:
\begin{equation}\label{eq:L_neutrinos}
\epsilon_{e^{-}\!e^{+}} \approx 8.69\times 10^{30}\left(\frac{k_B T}{1\,{\rm MeV}}\right)^9\,\, {\rm MeV}\,{\rm cm}^{-3}\,{\rm s}^{-1},
\end{equation}
where $k_B$ is the Boltzmann constant.

The accretion zone is characterized by a~temperature gradient with a~typical scale height \mbox{$\Delta r_{\rm ER} = T/\nabla T \approx 0.7~R_{\rm NS}$}. Owing to the aforementioned strong dependence of the neutrino emission on temperature, most of the neutrinos are emitted from a~spherical shell around the NS of thickness
\begin{equation}
\Delta r_{\nu} = \frac{\epsilon_{e^{-}\!e^{+}}}{\nabla \epsilon_{e^{-}\!e^{+}}} = \frac{\Delta r_{\rm ER}}{9} \approx 0.08R_{\rm NS}.
\label{neutrinoshell}
\end{equation}

Equations~(\ref{eq:Tacc}) and (\ref{eq:L_neutrinos}) imply the neutrino emissivity satisfies $\epsilon_{e^{-}\!e^{+}} \propto \dot{M}^{9/4}_{\rm acc}$ as we had anticipated. These conditions lead to the neutrinos to be efficient in balancing the gravitational potential energy gain, allowing the hypercritical accretion rates. The effective accretion onto the NS can be estimated as: 
\begin{equation}\label{eq:Mdoteff}
\dot{M}_{\rm eff} \approx \Delta M_{\nu} \frac{L_{\nu}}{E_g},
\end{equation}
where $\Delta M_{\nu}$, $L_{\nu}$ are, respectively, the mass and neutrino luminosity in the emission region, and~$E_g=(1/2) G M_{\rm NS} \Delta M_{\nu}/(R_{\nu}+\Delta r_{\nu})$ is half the gravitational potential energy gained by the material falling from infinity to the $R_{\rm NS}+\Delta r_{\nu}$. The neutrino luminosity is
\begin{equation}
L_{\nu}\approx 4\pi R_{\rm NS}^2\Delta r_{\nu} \epsilon_{e^{-}\!e^{+}},
\label{eq:neutrinoluminosity}
\end{equation}
with $\epsilon_{e^{-}\!e^{+}}$ being the neutrino emissivity in Equation~(\ref{eq:L_neutrinos}). For $M_{\rm NS}=2~M_\odot$ and temperatures $1$--10~MeV, the Equations~(\ref{eq:Mdoteff}) and (\ref{eq:neutrinoluminosity}) result $\dot{M}_{\rm eff} \approx 10^{-10}$--$10^{-1}~M_\odot$~s$^{-1}$ and $L_{\nu} \approx 10^{48}$--$10^{57}$~MeV~s$^{-1}$.

\begin{figure}[H]
\centering
\includegraphics[width=0.8\hsize,clip]{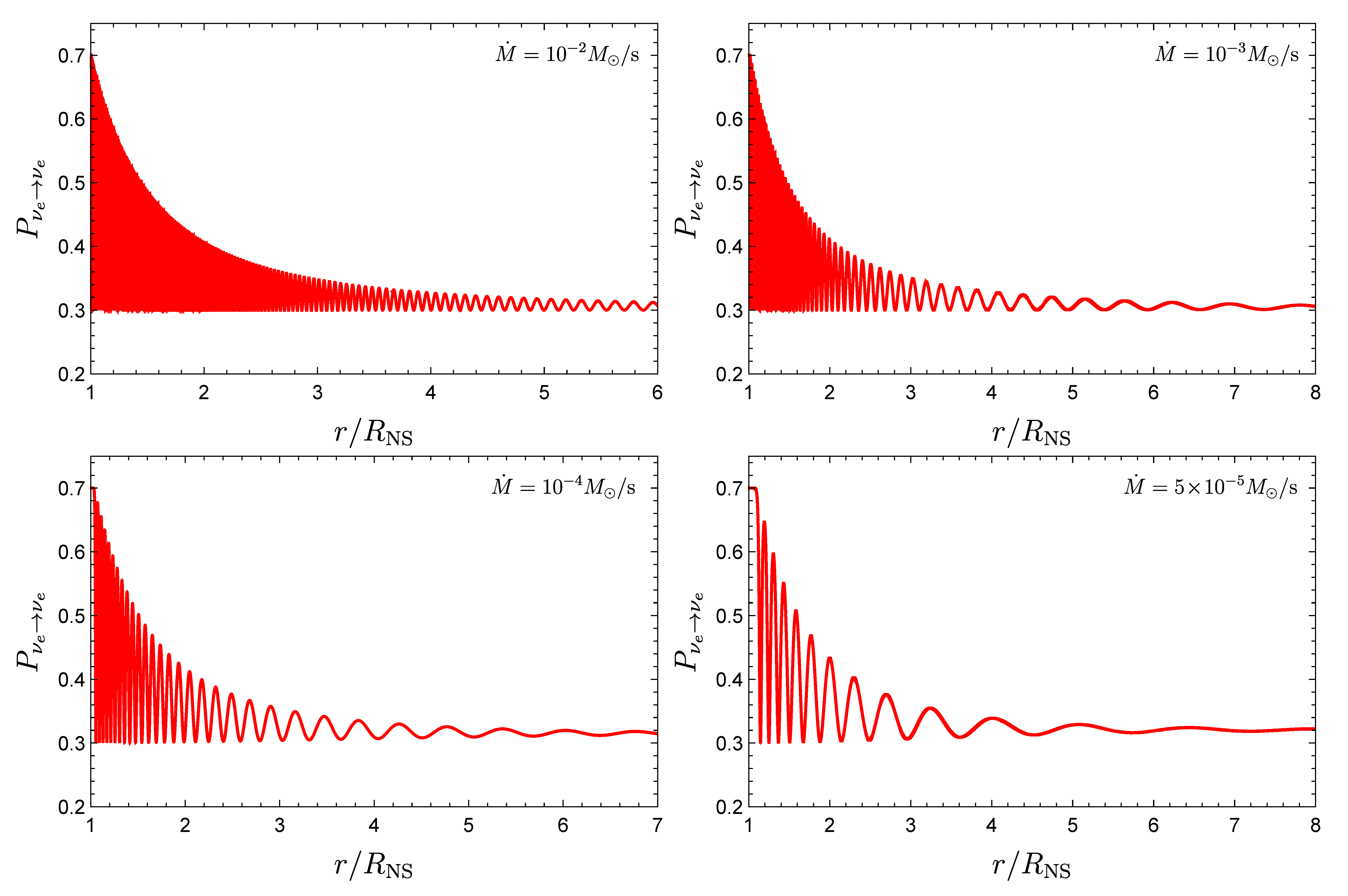}
\caption{Neutrino flavor evolution in the case of the neutrino-mass inverted hierarchy {(taken from Figure 4 in~\cite{2018ApJ...852..120B})}. The electron-neutrino survival probability is shown as a~function of the radial distance from the NS surface. The curves for the electron antineutrino overlap the ones for electron-neutrinos.} 
\label{fig:singleangle}
\end{figure}

Therefore, the neutrino emission can reach luminosities of up to $10^{57}$~MeV~s$^{-1}$, mean neutrino energies $20$--$30$~MeV, and neutrino densities $10^{31}$~cm$^{-3}$. Along their path from the vicinity of the NS surface outward, such neutrinos experience flavor transformations dictated by the neutrino to electron density ratio. We have determined in~\cite{2018ApJ...852..120B} the neutrino and electron on the accretion zone and use them to compute the neutrino flavor evolution. For normal and inverted neutrino-mass hierarchies and within the two-flavor formalism ($\nu_{e}\nu_{x}$), we estimated the final electronic and non-electronic neutrino content after two oscillation processes: (1) neutrino collective effects due to neutrino self-interactions where the neutrino density dominates and, (2) the Mikheyev-Smirnov-Wolfenstein (MSW) effect, where the electron density dominates. We find that the final neutrino content is composed by $\sim$55\% ($\sim$62\%) of electronic neutrinos, i.e., $\nu_{e}+\bar{\nu}_{e}$, for the normal (inverted) neutrino-mass hierarchy (see~Figure~\ref{fig:singleangle}). This is a~first step toward the characterization of a~novel source of astrophysical MeV-neutrinos in addition to core-collapse SNe. We refer the reader to~\cite{2018ApJ...852..120B} for additional details of the flavor-oscillations as well as the final neutrino spectra after such a~process.

%%%%%%%%%%%%%%%%%%%%%%%%%%%%%%%%%%%%%%%%%%%%%%%%%%%%%%%%%%%%%%%%
\subsection{Accretion Luminosity}
%%%%%%%%%%%%%%%%%%%%%%%%%%%%%%%%%%%%%%%%%%%%%%%%%%%%%%%%%%%%%%%%

The energy release in a~time-interval $dt$, when an~amount of mass $dM_b$ with  angular momentum $l \dot{M}_b$ is accreted, is~\cite{2016ApJ...833..107B}:
\begin{equation}
L_{\rm acc}= (\dot{M}_b - \dot{M}_{\rm NS})c^2 =\dot{M}_b c^2 \left[1-\left(\frac{\partial M_{\rm NS}}{\partial J_{\rm NS}}\right)_{M_b}\,l -\left(\frac{\partial M_{\rm NS}}{\partial M_b}\right)_{J_{\rm NS}}\right].
\label{eq:DiskLuminosity}
\end{equation}

This is the amount of gravitational energy gained by the matter by infalling to the NS surface that is not spent in NS gravitational binding energy. The total energy release in the time interval from $t$ to $t+dt$,
\begin{equation}
\Delta E_{\rm acc} \equiv \int L_{\rm acc}dt,
\end{equation}
is given by the NS binding energy difference between its initial and final state. The typical luminosity is $L_{\rm acc}\approx \Delta E_{\rm acc}/\Delta t_{\rm acc}$, where $\Delta t_{\rm acc}$ is the duration of the accretion process.

The value of $\Delta t_{\rm acc}$ is approximately given by the flow time of the slowest layers of the SN ejecta to the NS companion position. If we denote the velocity of these layers by $v_{\rm inner}$, we have $\Delta t_{\rm acc}\sim a/v_{\rm inner}$, where $a$ is the binary separation. For $a\sim 10^{11}$~cm and $v_{\rm inner}\sim 10^8$~cm~s$^{-1}$, $\Delta t_{\rm acc}\sim 10^3$~s. For shorter separations, e.g.,~$a\sim 10^{10}$~cm ($P\sim 5$~min), $\Delta t_{\rm acc}\sim 10^2$~s. For a~binary with $P=5$~min, the NS accretes $\approx$1~$M_\odot$ in $\Delta t_{\rm acc}\approx 100$~s. From Equation~(\ref{eq:MbMnsjns}) one obtains that the binding energy difference of a~$2~M_\odot$ and a~$3~M_\odot$ NS, is $\Delta E_{\rm acc}\approx 13/200 (3^2-2^2)~M_\odot c^2\approx 0.32~M_\odot c^2$. This leads to $L_{\rm acc}\approx 3\times 10^{-3}~M_\odot c^2 \approx 0.1\, \dot{M_b} c^2$. The accretion power can be as high as $L_{\rm acc}\sim 0.1 \dot{M_b} c^2\sim 10^{47}$--$10^{51}$~erg~s$^{-1}$ for accretion rates in the range $\dot{M_b}\sim 10^{-6}$--$10^{-2}~M_\odot$~s$^{-1}$.

\section{New 3D SPH Simulations}\label{sec:4}
%%%%%%%%%%%%%%%%%%%%%%%%%%%%%%%%%%%%%%%%%%%%%%%%%%%%%%%%%%%%
%%%%%%%%%%%%%%%%%%%%%%%%%%%%%%%%%%%%%%%%%%%%%%%%%%%%%%%%%%%%

We have recently presented in~\cite{2019ApJ...871...14B} new, 3D hydrodynamic simulations of the IGC scenario by adapting the SPH code developed at Los Alamos, \emph{SNSPH}~\cite{2006ApJ...643..292F}, which has been tested and applied in a~variety of astrophysical situations~\cite{2002ApJ...574L..65F,2006ApJ...640..891Y,2008ASPC..391..221D,2017ApJ...846L..15B}.

\begin{figure}[H]
  \centering
  \includegraphics[width=\hsize,clip]{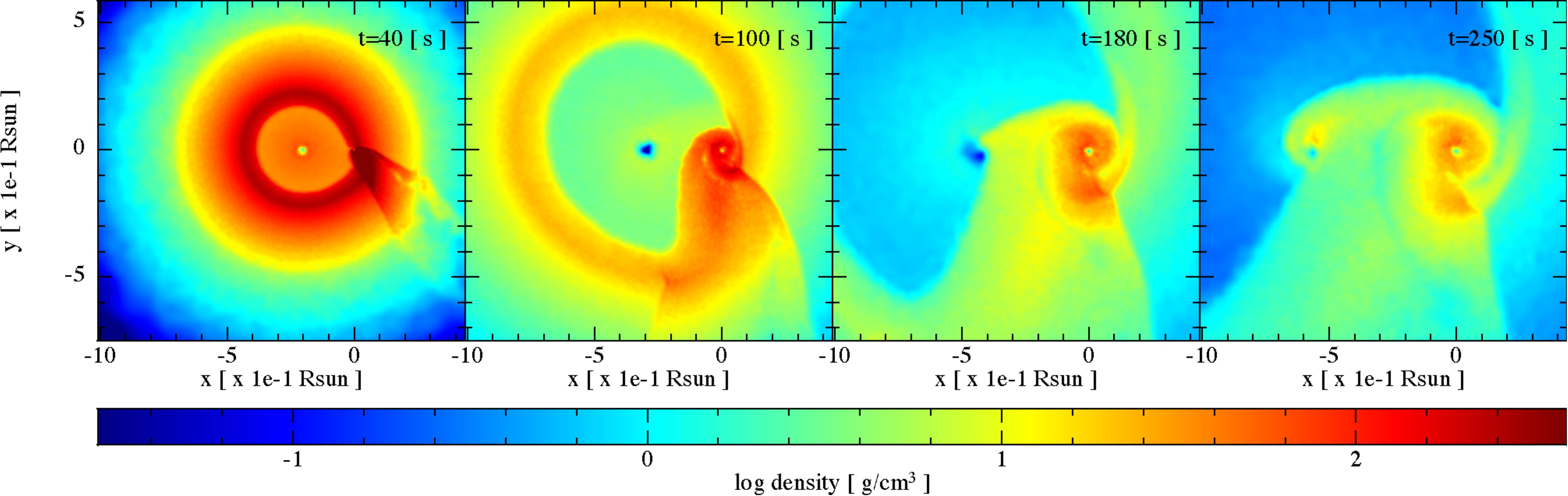} \includegraphics[width=\hsize,clip]{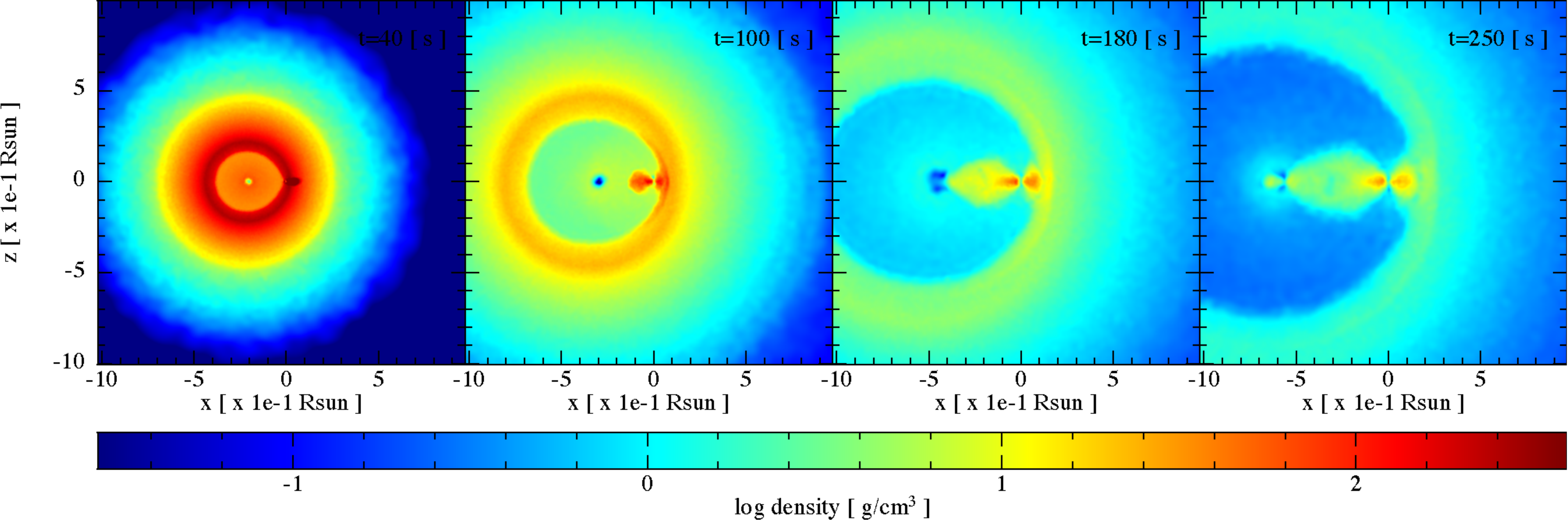}
 \caption{Snapshots of the 3D SPH simulations of the IGC scenario {(taken from Figure~2 in~\cite{2019ApJ...871...14B})}. The~initial binary system is formed by a~CO$_{\rm core}$ of mass $\approx 6.85~M_\odot$, from a~ZAMS progenitor star of $25~M_\odot$, and a~$2~M_\odot$ NS with an~initial orbital period of approximately $5$~min. The upper panel shows the mass density on the equatorial (orbital) plane, at different times of the simulation. The time $t=0$ is set in our simulations at the moment of the SN shock breakout. The lower panel shows the plane orthogonal to the orbital one. The reference system has been rotated and translated for the \emph{x}-axis to be along the line joining the $\nu$NS and the NS centers, and its origin is at the NS position.}
  \label{fig:3DSPH}
\end{figure}

The time $t=0$ of the simulation is set as the time at which the SN shock breaks out the CO$_{\rm core}$ external radius. We calculate the accretion rate both onto the NS companion and onto the $\nu$NS (via fallback), and calculate the evolution of other binary parameters such as the orbital separation, eccentricity, etc. Figure~\ref{fig:3DSPH} shows an~example of simulation for a~binary system composed of a~CO$_{\rm core}$ of mass $\approx$6.85~$M_\odot$, the end stage of a~ZAMS progenitor star of $M_{\rm zams}=25~M_\odot$, and a~$2~M_\odot$ NS companion. The initial orbital period is $\approx$5~min.

\begin{figure}[H]
  \centering
  \includegraphics[width=0.48\hsize]{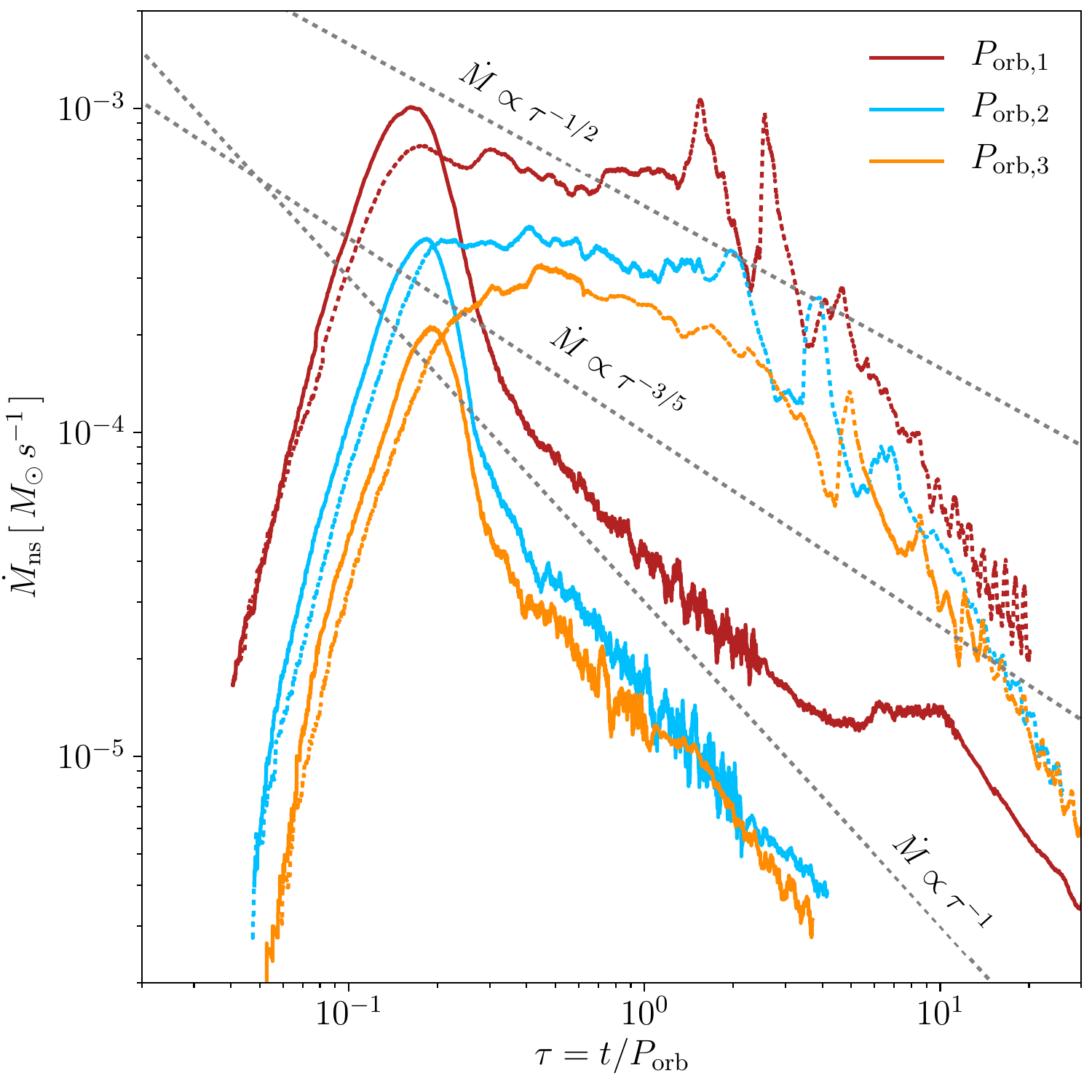}\includegraphics[width=0.48\hsize]{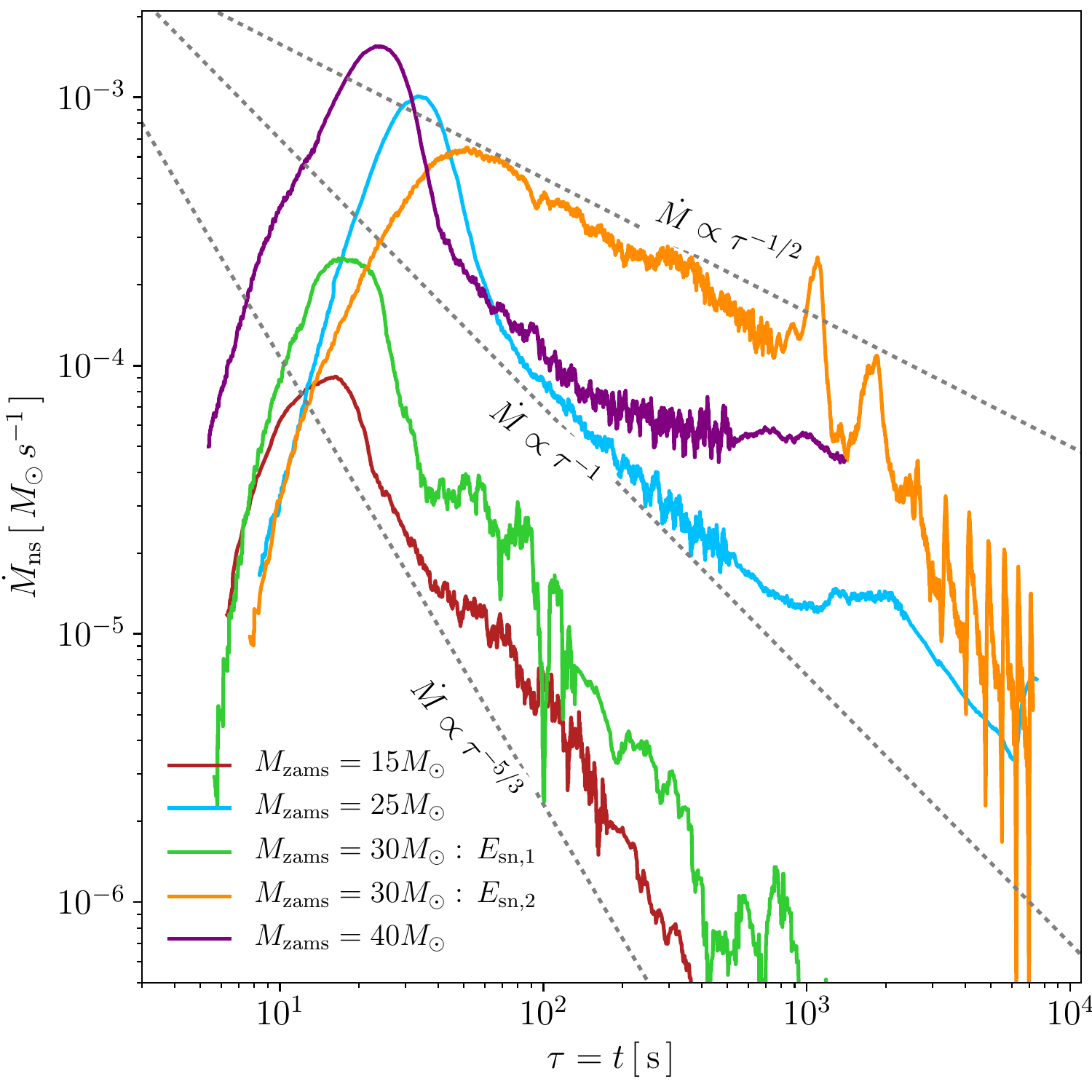}
  \caption{(\textbf{a}) Mass-accretion rate onto the NS companion in the IGC scenario {(taken from Figure~9 in~\cite{2019ApJ...871...14B})}. Different colors correspond to different initial orbital periods: $P_{\rm orb,1}=4.8$~min (red line), $P_{\rm orb,1}=8.1$~min (blue line), $P_{\rm orb,1}=11.8$~min (orange line). The other parameters that characterize the initial binary system are the same as in Figure~\ref{fig:3DSPH}. The solid lines correspond to a~SN energy of $1.57\times 10^{51}$~erg, while the dotted ones correspond to a~lower SN energy of $6.5 \times 10^{50}$~erg. It can be seen that the mass-accretion rate scales with the binary orbital period. (\textbf{b}) Mass-accretion rate on the NS companion for all the CO$_{\rm core}$ progenitors {(see Table~1 and Figure~13 in~\cite{2019ApJ...871...14B})}. The NS companion has an~initial mass of $2\,M_\odot$ and the orbital period is close to the minimum period that the system can have in order that there is no Roche-lobe overflow before the collapse of the CO$_{\rm core}$: $6.5$~min, $4.8$~min, $6.0$~min and $4.4$~min for the $M_{\rm zams}=15 M_\odot$, $25 M_\odot$, $30M_\odot$ and $40M_\odot$ progenitors, respectively.}
  \label{fig:Mdots}
\end{figure}

The accretion rate onto both stars was estimated from the flux of SPH particles falling, per unit time, into the Bondi-Hoyle accretion region of the NS (see~Figure~\ref{fig:Mdots}). It is confirmed that the accretion onto the NS companion occurs from a~disk-like structure formed by the particles that circularize before being accreted; see vortexes in the upper panel of Figure~\ref{fig:3DSPH} and the disk structure is clearly seen in the lower panel.

Several binary parameters were explored thanks to the new code. We performed simulations changing the CO$_{\rm core}$ mass, the NS companion mass, the orbital period, the SN explosion energy (so~the SN kinetic energy or velocity). We also explored intrinsically asymmetric SN explosion. We~checked if the $\nu$NS and/or the NS companion reach the mass-shedding (Keplerian) limit or the secular axisymmetric instability, i.e.,~the critical mass. The NS can also become just a~more massive, fast~rotating, stable NS when the accretion is moderate. All this was done for various NS nuclear equations of state (NL3, TM1 and GM1).

We followed the orbital evolution up to the instant when most of the ejecta has abandoned the system to determine if the system remains bound or becomes unbound by the explosion. We thus assessed the CO$_{\rm core}$-NS parameters leading to the formation of $\nu$NS-NS (from XRFs) or $\nu$NS-BH (from~BdHNe) binaries. The first proof that BdHNe remain bound leading to $\nu$NS-BH binaries was presented in~\cite{2015PhRvL.115w1102F} (see next section).

%%%%%%%%%%%%%%%%%%%%%%%%%%%%%%%%%%%%%%%%%%%%%%%%%%%%%%%%%%%%%%%%
%%%%%%%%%%%%%%%%%%%%%%%%%%%%%%%%%%%%%%%%%%%%%%%%%%%%%%%%%%%%%%%%
\section{Consequences on GRB Data Analysis and Interpretation}\label{sec:5}
%%%%%%%%%%%%%%%%%%%%%%%%%%%%%%%%%%%%%%%%%%%%%%%%%%%%%%%%%%%%%%%%
%%%%%%%%%%%%%%%%%%%%%%%%%%%%%%%%%%%%%%%%%%%%%%%%%%%%%%%%%%%%%%%%

{
In a~few seconds a~BdHN shows different physical processes that lead to a~specific sequence of observables at different times and at different wavelengths. Starting with the at-times-observed X-ray precursors, to the Gamma-ray prompt emission, to the GeV emission, to the early and late X-ray afterglow in which, respectively, are observed flares and a~distinct power-law luminosity. 
}

%%%%%%%%%%%%%%%%%%%%%%%%%%%%%%%%%%%%%%%%%%%%%%%%%%%%%%%%
\subsection{X-ray Precursor}

{X-ray precursors can comprise the presence of both the SN shock breakout as well as the hypercritical accretion onto the NS companion until it reaches the critical mass. These processes have been identified in~\cite{2012A&A...548L...5I,2014ApJ...793L..36F,2016ApJ...833..107B,2019ApJ...874...39W}.}

{The conversion of the SN shockwave kinetic energy (see~\cite{1996snih.book.....A} for details on the SN physics) into electromagnetic energy imply that about $10^{50}$~erg can be emitted.}

{Once it reached the NS companion, the ejecta induced a~hypercritical accretion onto the NS at a~rate $\sim$10$^{-3} M_\odot$~s$^{-1}$ for an~assumed orbital separation of few $10^{10}$~cm. As we have recalled (see~Figure~\ref{fig:Ep1GRB090618} in Section~\ref{sec:3}), the accretion process triggers the expansion of thermal convective bubbles on top of the NS owing to the Rayleigh-Taylor instability~\cite{2012A&A...548L...5I,2014ApJ...793L..36F,2016ApJ...833..107B,2019ApJ...874...39W}. }

{It is of special interest to refer the reader to the results presented in~\cite{2019ApJ...874...39W} on GRB 180728A, a~BdHN II. It has been identified in the precursor of this GRB, for the first time, the presence of both the emergence of the SN shockwave as well as the hypercritical accretion process. From this the binary parameters have been extracted and further confirmed by the analysis of the prompt and the afterglow~emission.}

%%%%%%%%%%%%%%%%%%%%%%%%%%%%%%%%%%%%%%%%%%%%%%%%%%%%%%%%%%%%%%%%
\subsection{{GRB Prompt Emission}}
%%%%%%%%%%%%%%%%%%%%%%%%%%%%%%%%%%%%%%%%%%%%%%%%%%%%%%%%%%%%%%%%

{A BdHN I leaves as a~remnant a~$\nu$NS-BH binary surrounded by the asymmetric SN ejecta (see~Figure~\ref{fig:profiles} and~\cite{2016ApJ...833..107B,2019ApJ...871...14B}). The asymmetric ejecta includes a~``cavity'' of $\sim$10$^{11}$~cm of very low-density matter around the newborn BH. The hydrodynamics inside such a~low-density cavity have been recently studied by numerical simulations in~\cite{2019arXiv190403163R}.}

\begin{figure}[H]
    \centering
    \includegraphics[width=0.8\hsize,clip]{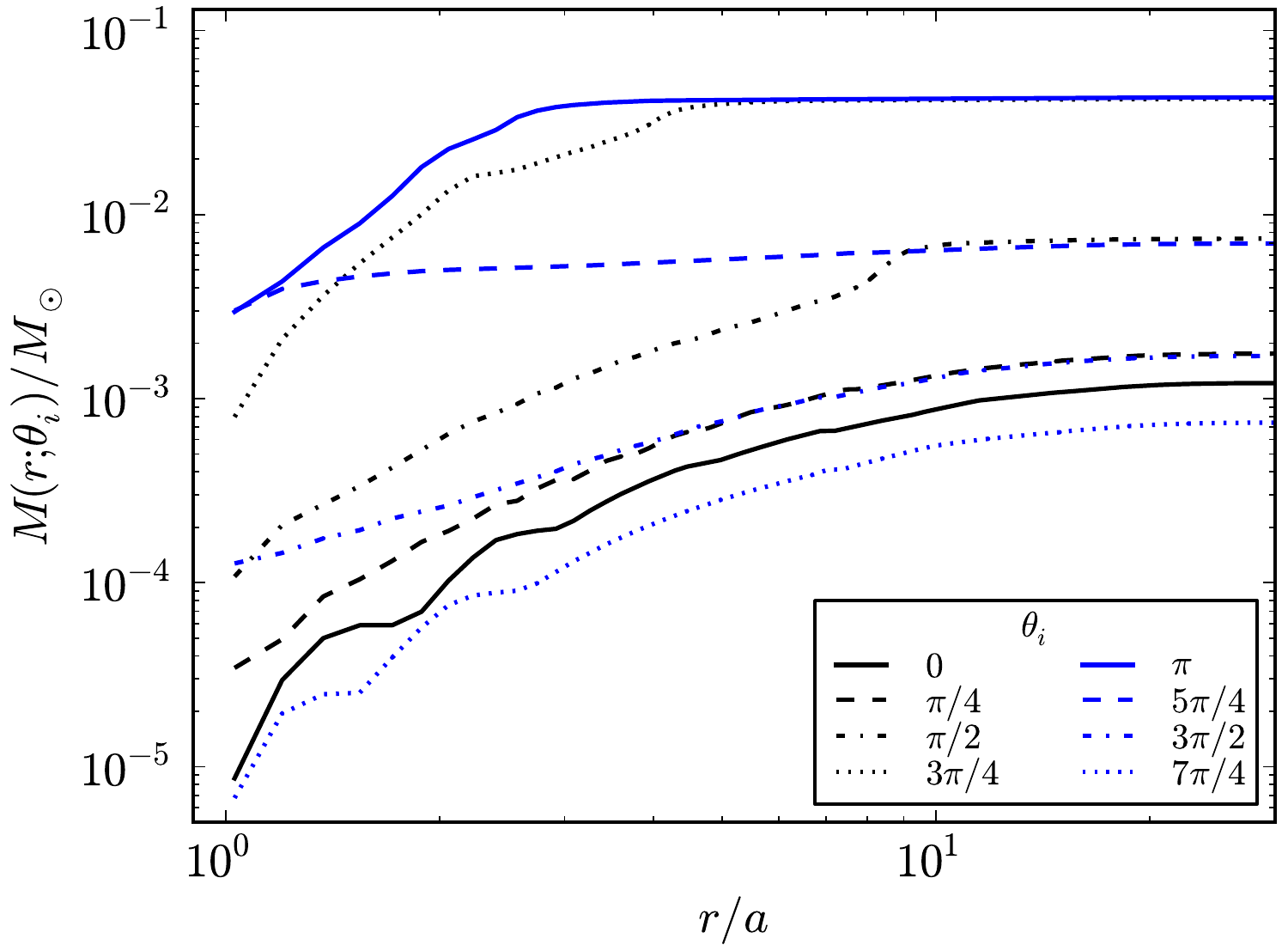}
    \caption{Cumulative radial mass profiles of the SN ejecta enclosed within a~cone of $5^\circ$ of semi-aperture angle with vertex at the BH position {(taken from Figure~35 in~\cite{2018ApJ...852...53R}). These profiles have been extracted from the simulations at the time of BH formation}. The binary parameters are: the initial mass of the NS companion is $2.0~M_\odot$; the CO$_{\rm core}$ leading to an~ejecta mass of $7.94~M_\odot$, and the orbital period is $P\approx 5$~min, namely a~binary separation $a\approx 1.5\times 10^{10}$~cm.}
    \label{fig:profiles2}
\end{figure}

{The asymmetric character acquired by the SN ejecta implies that the $e^+e-$ plasma, expanding from the BH site in all directions with equal initial conditions, experiences a~different dynamics along different directions. The reason for this is that the $e^+e-$ plasma engulfs different amounts of baryonic mass (see~Figure~\ref{fig:profiles2}). This leads to observable signatures as a~function of the viewing angle.}

{The newborn Kerr BH, surrounded by ejecta and immersed in a~test magnetic field (likely the one left by the magnetized, collapsed NS), represents what we have called the \emph{inner engine} of the high-energy emission~\cite{2018arXiv181101839R,2018arXiv181200354R,2019arXiv190404162R,2019arXiv190403163R}. The rotating BH, of mass $M$ and angular momentum $J$, in the presence of the magnetic field $B_0$, induces an~electromagnetic field described by the Wald solution~\cite{1974PhRvD..10.1680W}. }

{The induced electric field at the BH horizon $r_+ = M (1+\sqrt{1-\alpha^2})$ is~\cite{2018arXiv181101839R,2018arXiv181200354R}
\begin{equation}\label{eq:Eh}
    E_{r_+}\approx  \frac{1}{2}\alpha B_0= 6.5\times 10^{15}\cdot \alpha \left(\frac{B_0}{B_c}\right) \quad \frac{{\rm V}}{\rm cm},
\end{equation}
where $\alpha = J/M^2$ is the dimensionless angular momentum of the BH and $B_c=m_e^2 c^3/(e \hbar)\approx 4.4\times 10^{13}$~G. This field acquires values over the critical one, $E_c = m_e^2 c^3/(e\hbar)$ if the following conditions are~verified:
\begin{equation}\label{eq:conditions}
    \alpha (B_0/B_c) \geq 2,\qquad B_0/B_c\geq 2,
\end{equation}
where the second condition comes from the constraint that a~rotating BH must satisfy: $\alpha\leq 1$. The~above huge value of the electric field (\ref{eq:Eh}) guarantees the production of the $e^+e^-$ pair plasma around the newborn BH via the quantum electrodynamics (QED) process of vacuum polarization~\cite{2010PhR...487....1R}.}

In the direction pointing from the CO$_{\rm core}$ to the accreting NS outwards and lying on the orbital plane, { the aforementioned cavity represents a~region of low baryonic contamination~\cite{2019arXiv190403163R,2019arXiv190404162R}. The $e^+e-$ plasma can then self-accelerate to Lorentz factors $\Gamma \sim 10^2$--$10^3$ reaching transparency and impacting on the CBM filaments as described in~\cite{1998A&A...338L..87P,1999A&AS..138..511R,2000A&A...359..855R}. At transparency, MeV-photons are emitted which are observed in the ultrarelativistic prompt emission. This picture has been successfully applied and verified on plenty of GRBs, e.g.,~GRBs 050904, 080319B, 090227, 090618 and 101023~\cite{2012A&A...538A..58P,2012A&A...543A..10I,2012ApJ...756...16P,2013ApJ...763..125M}}. 

%%%%%%%%%%%%%%%%%%%%%%%%%%%%%%%%%%%%%%%%%%%%%%%%%%%%%%%%%%%%%%%
\subsection{{Early X-ray Afterglow: Flares}}
%%%%%%%%%%%%%%%%%%%%%%%%%%%%%%%%%%%%%%%%%%%%%%%%%%%%%%%%%%%%%%%

It was recently addressed in~\cite{2018ApJ...852...53R} the role of X-ray flares as a~powerful tool to differentiate the BdHN model from the ``collapsar-fireball'' model~\cite{1993ApJ...405..273W}. 

First, it is known that the GRB prompt emission shows Gamma-ray spikes occurring at $10^{15}$--$10^{17}$~cm from the source and have Lorentz factor $\Gamma \sim 10^2$--$10^3$.

Second, the thermal emission observed in the X-ray flares of the early (rest-frame time $t\sim 10^2$~s) afterglow of BdHNe, implies occurrence radii $\sim$10$^{12}$~cm expanding at mildly-relativistic velocity, e.g.,~$\Gamma\lesssim 4$~\cite{2018ApJ...852...53R} (see below). {The latter observational fact evidences that the X-ray afterglow is powered by a~mildly-relativistic emitter. These model-independent observations contrast with the assumption of an~ultrarelativistic expansion starting from the GRB prompt emission and extending to the afterglow. Such a~``traditional'' approach to GRBs has been adopted in a~vast number of articles over decades as it is summarized in review articles (see, e.g.,~\cite{1999PhR...314..575P,2004RvMP...76.1143P,2002ARA26A..40..137M,2006RPPh...69.2259M,2014ARA&A..52...43B,2015PhR...561....1K}.}

In the other directions, the GRB $e^+e-$ plasma impacts the SN ejecta at approximately $10^{10}$~cm, evolves carrying a~large amount of baryons reaching transparency at radii $10^{12}$~cm with a~mildly $\Gamma\lesssim 4$. The theoretical description and the consequent numerical simulation have been addressed in~\cite{2018ApJ...852...53R}. 

Such a~mildly-relativistic photospheric emission is experimentally demonstrated by the thermal radiation observed in the early X-ray afterglow and in the X-ray flares~\cite{2012MNRAS.427.2950S,2018MmSAI..89..293W}. For instance, in the early hundreds of seconds, GRB 090618 is found to have a~velocity of $\beta \sim 0.8$~\cite{2014A&A...565L..10R,2011MNRAS.416.2078P}, GRB 081008 has a~velocity $\beta \sim 0.9$~\cite{2018ApJ...852...53R}, and GRB 130427A has a~velocity of $\beta \sim 0.9$ as well~\cite{2015ApJ...798...10R,2015ARep...59..667W,2018ApJ...869..101R}. We emphasize that the mildly-relativistic photo-sphere velocity is derived from the data in a~model-independent way, summarising from~\cite{2018ApJ...852...53R}:
\begin{multline}
\frac{\beta^5}{4 [ \ln (1+ \beta) - (1-\beta) \beta]^2} \left(\frac{1+\beta}{1-\beta}\right)^{1/2} = \frac{D_L(z)}{1+z} \frac{1}{t_2-t_1} \left(\sqrt{\frac{F_\mathrm{bb,obs} (t_2)}{\sigma T_\mathrm{obs}^4(t_2)}} - \sqrt{\frac{F_\mathrm{bb,obs}(t_1)}{\sigma T_\mathrm{obs}^4(t_1)}}\right) ,
\label{betaNew}
\end{multline}

The left-hand side is a~function of velocity $\beta$, the right-hand side is only from observables, $D_L(z)$ is the luminosity distance and $z$ the cosmological redshift. From the observed thermal flux $F_\mathrm{bb,obs}$ and temperature  $T_\mathrm{obs}$ in two times $t_1$ and $t_2$, the velocity $\beta$ is obtained. This model-independent equation has been derived in a~fully relativistic way so it remains valid in the Newtonian non-relativistic regime. 

%The SN ejecta morphology also explains the presence or absence of GeV emission in BdHNe (see~\cite{2018arXiv180305476R} for details).

%

An additional, and very important prediction of this scenario, is that the injection of energy and momentum from the GRB plasma into the ejecta transforms the SN into an~HN (see~\cite{2018ApJ...869..151R} for the specific case of GRB 151027A).

%%%%%%%%%%%%%%%%%%%%%%%%%%%%%%%%%%%%%%%%%%%%%%%%%%%%%%%%%%%%%%%%
\subsection{{Late X-ray Afterglow}}
%%%%%%%%%%%%%%%%%%%%%%%%%%%%%%%%%%%%%%%%%%%%%%%%%%%%%%%%%%%%%%%%

{
We have shown in \citet{2018ApJ...869..101R} that the  synchrotron emission by relativistic electrons from the $\nu$NS, injected into the expanding magnetized HN ejecta, together with the $\nu$NS pulsar emission that extracts its rotational energy, power the X-ray afterglow. This includes the early part and the late power-law behavior. An exceptional by-product of this analysis is that it gives a~glimpse on the $\nu$NS magnetic field strength and structure (dipole+quadrupole).
}

{
Based on the above model~\cite{2018ApJ...869..101R}, GRB 130427A (a BdHN I) and GRB 180728A (a BdHN II) have been analyzed in~\cite{2019ApJ...874...39W}. The explanation of the afterglow data of GRB 130427A led to an~initial $1$~ms rotation period for the $\nu$NS. For GRB 180728A, a~slower spin of $2.5$~ms was there obtained. A simple analysis showed how this result is in agreement with the BdHN I and II nature of these GRBs. First, we recall that compact binary systems have likely synchronized components with the orbital period. Second, we can infer the orbital period from the analysis of the X-ray precursor and the prompt emission (see~\cite{2019ApJ...874...39W} for the procedure). Then, we can infer the CO$_{\rm core}$ rotation period too. Finally, assuming angular momentum conservation in the core-collapse SN process, we can estimate the rotation period of the $\nu$NS formed at the SN center. This method led to a~binary separation remarkably in agreement with the one inferred from the precursor and the prompt emission, demonstrating the self-consistency of this scenario~\cite{2019ApJ...874...39W}.
}

%%%%%%%%%%%%%%%%%%%%%%%%%%%%%%%%%%%%%%%%%%%%%%%%%%%%%%%%%%%%%%%%
\subsection{{High-Energy GeV Emission}}
%%%%%%%%%%%%%%%%%%%%%%%%%%%%%%%%%%%%%%%%%%%%%%%%%%%%%%%%%%%%%%%%

{
We turn back again to the already introduced \emph{inner engine}. The joint action of rotation and magnetic field induces an~electric potential~\cite{2018arXiv181101839R, 2018arXiv181200354R}
\begin{equation}\label{eq:deltaphi}
    \Delta \phi = - \int_\infty^{r_+} E dr = E_{r_+} r_+ =9.7\times 10^{20}\cdot \alpha \left(\frac{B_0}{B_c}\right) \left(\frac{M}{M_\odot}\right) (1+\sqrt{1-\alpha^2})\quad\frac{{\rm V}}{e},
\end{equation}
capable to accelerate protons to ultrarelativistic velocities and energies up to $\epsilon_p = e \Delta\phi\approx 10^{21}$~eV.
}

{
Along the rotation axis, there are no radiation losses and so the \emph{inner engine} leads to UHECRs. In~the off-polar directions, the protons radiate synchrotron photons, e.g., at GeV and TeV energies.
}

{
In~\cite{2018arXiv181101839R} it has been estimated that the available electrostatic energy to accelerate protons is
\begin{equation}
{\cal E} = \frac{1}{2} E_{r_+}^2 r_+^3 \approx 7.5\times 10^{41}\cdot\alpha^2 \left(\frac{B_0}{B_c}\right)^2 \left(\frac{M}{M_\odot}\right)^3 (1+\sqrt{1-\alpha^2})^3\quad{\rm erg},
\end{equation}
so the number of protons that the \emph{inner engine} can accelerate is
\begin{equation}\label{eq:Np}
    N_p = \frac{{\cal E}}{\epsilon_p} \approx 4.8\times 10^{32} \alpha \left(\frac{B_0}{B_c}\right) \left(\frac{M}{M_\odot}\right)^2 (1+\sqrt{1-\alpha^2})^2.
\end{equation}    
}

{
The timescale of the first elementary process is dictated by the acceleration time, i.e.,:
\begin{equation}\label{eq:timescale}
    \Delta t_{\rm el} = \frac{\Delta \phi}{E_{r_+} c} = \frac{r_+}{c} \approx 4.9\times 10^{-6}\,\left(\frac{M}{M_\odot}\right) (1+\sqrt{1-\alpha^2})\quad {\rm s}.
\end{equation}
so the emission power of the \emph{inner engine} is approximately:
\begin{equation}\label{eq:power}
    \frac{d{\cal E}}{dt} \approx \frac{{\cal E}}{\Delta t_{\rm el}} = 1.5\times 10^{47}\cdot\alpha^2 \left(\frac{B_0}{B_c}\right)^2 \left(\frac{M}{M_\odot}\right)^2 (1+\sqrt{1-\alpha^2})^2\quad{\rm erg}\cdot {\rm s}^{-1}.
\end{equation}
}

{
The timescale of the subsequent processes depends crucially on the time required to rebuild the electric field.  It has been shown that this condition implies an~essential role of the density profile of the ionic matter surrounding the BH and its evolution with time~\cite{2018arXiv181200354R,2019arXiv190404162R}. %meaning retained?
}

{
For a~BH mass of the order of the NS critical mass, say $M\sim 3~M_\odot$, a~BH spin parameter $\alpha\sim 0.3$, and a~strength of the magnetic field $B_0 \sim 10^{14}$~G, the above numbers are in agreement with the observed GeV emission data. See, for instance, in~\cite{2018arXiv181200354R} and~\cite{2019arXiv190404162R}, respectively, the details of the analysis of GRB 130427A and GRB 190114C. We refer to~\cite{2018arXiv181101839R,2018arXiv181200354R} for details on the synchrotron emission of the accelerated protons in the above magnetic field. 
}

%%%%%%%%%%%%%%%%%%%%%%%%%%%%%%%%%%%%%%%%%%%%%%%%%%%%%%%%%%%%%%%%
\subsection{{Additional Considerations}}
%%%%%%%%%%%%%%%%%%%%%%%%%%%%%%%%%%%%%%%%%%%%%%%%%%%%%%%%%%%%%%%%

The strong dependence of $P_{\rm max}$ on the initial mass of the NS companion opens the interesting possibility of producing XRFs and BdHNe from binaries with similar short (e.g.,~$P\sim$ few minutes) orbital periods and CO$_{\rm core}$ properties: while a~system with a~massive (e.g.,~$\gtrsim$2~$M_\odot$) NS companion would lead to a~BdHN, a~system with a~lighter (e.g.,~$\lesssim$1.4~$M_\odot$) NS companion would lead to an~XRF. This predicts systems with a~similar initial SN, leading to a~similar $\nu$NS, but with different GRB prompt and afterglow emission. Given that the GRB energetics are different, the final SN kinetic energy should also be different being that it is larger for the BdHNe.
{This has been clearly shown by specific examples in~\cite{2019ApJ...874...39W}.}

There are also additional novel features unveiled by the new 3D SPH simulations which can be observable in GRB light-curves and spectra, e.g.,:

\begin{enumerate} [leftmargin=2.3em,labelsep=4mm]
\item[(1)] the hypercritical accretion occurs not only on the NS companion but also on the $\nu$NS and with a~comparable rate.

\item[(2)] This implies that BdHNe might be also be able to form, in special cases, BH-BH binaries. Since the system remains bound the binary will quickly merge by emitting gravitational waves. Clearly, no electromagnetic emission is expected from these mergers. However, the typically large cosmological distances of BdHNe would make it extremely difficult to detect their gravitational waves e.g., by LIGO/Virgo.

\item[(3)] Relatively weak SN explosions produce a~long-lived hypercritical accretion process leading and enhance, at late times, the accretion onto the $\nu$NS. The revival of the accretion process at late times is a~unique feature of our binary and does not occur for single SNe, namely in the absence of the NS companion. This feature increases the probability of detection of weak SNe by X-ray detectors via the accretion phase in an~XRF/BdHN.

\item[(4)] For asymmetric SN explosions the accretion rate shows a~quasi-periodic behavior that might be detected by X-rays instruments, possibly allowing a~test of the binary nature and the identification of the orbital period of the progenitor.
\end{enumerate}

\section{Post-Explosion Orbits and Formation of NS-BH Binaries}\label{sec:6}
%%%%%%%%%%%%%%%%%%%%%%%%%%%%%%%%%%%%%%%%%%%%%%%%%%%%%%%%%%%%
%%%%%%%%%%%%%%%%%%%%%%%%%%%%%%%%%%%%%%%%%%%%%%%%%%%%%%%%%%%%

The SN explosion leaves as a~central remnant a~$\nu$NS and the induced gravitational collapse of the NS companion leads to BH formation. Therefore, BdHNe potentially leads to $\nu$NS-BH binaries, providing the binary keeps bound. This question was analyzed via numerical simulations in~\cite{2015PhRvL.115w1102F}.

Typical binaries become unbound during an~SN explosion because of mass loss and the momentum imparted (kick) to the $\nu$NS by the explosion. A classical astrophysical result shows that, assuming the explosion as instantaneous (sudden mass loss approximation), disruption occurs if half of the binary mass is lost. For this reason the fraction of massive binaries that can produce double compact-object binaries is usually found to be very low (e.g.,~$\sim$0.001--1\%)~\cite{1999ApJ...526..152F,2012ApJ...759...52D,2014LRR....17....3P}.

Assuming instantaneous mass loss, the post-explosion semi-major axis is~\cite{1983ApJ...267..322H}: 
\begin{equation}
\frac{a}{a_0}=\frac{M_0 - \Delta M}{M_0 - 2 a_0 \Delta M/r},
\end{equation}
where $a_0$ and $a$ are the initial and final semi-major axes respectively, $M_0$ is the (initial) binary mass, $\Delta M$ is the change of mass (in this case the amount of mass loss), and $r$ is the orbital separation before the explosion. For circular orbits, the system is unbound if it loses half of its mass. For the very tight BdHNe, however, additional effects have to be taken into account to determine the fate of the binary.

The shock front in an~SN moves at roughly $10^4$~km~s$^{-1}$, but the denser, lower-velocity ejecta, can move at velocities as low as $10^2$--$10^3$~km~s$^{-1}$~\cite{2014ApJ...793L..36F}. This implies that the SN ejecta overcomes an~NS companion in a~time 10--1000~s. For wide binaries this time is a~small fraction of the orbital period and the ``instantaneous'' mass-loss assumption is perfectly valid. BdHNe have instead orbital periods as short as 100--1000~s, hence the instantaneous mass-loss approximation breaks down.

\begin{figure}[H]
    \centering
    \includegraphics[width=0.48\hsize,clip]{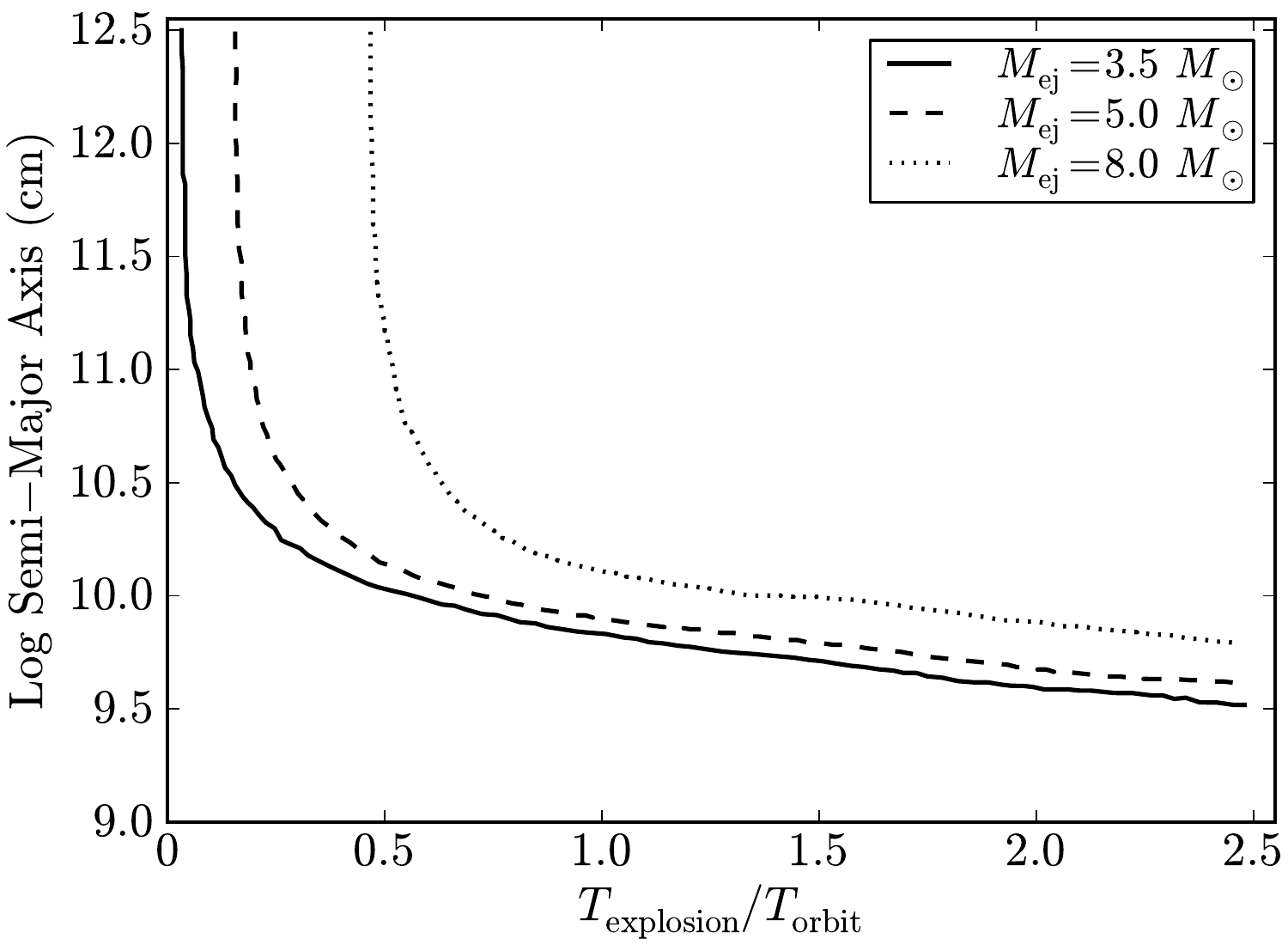}\includegraphics[width=0.48\hsize,clip]{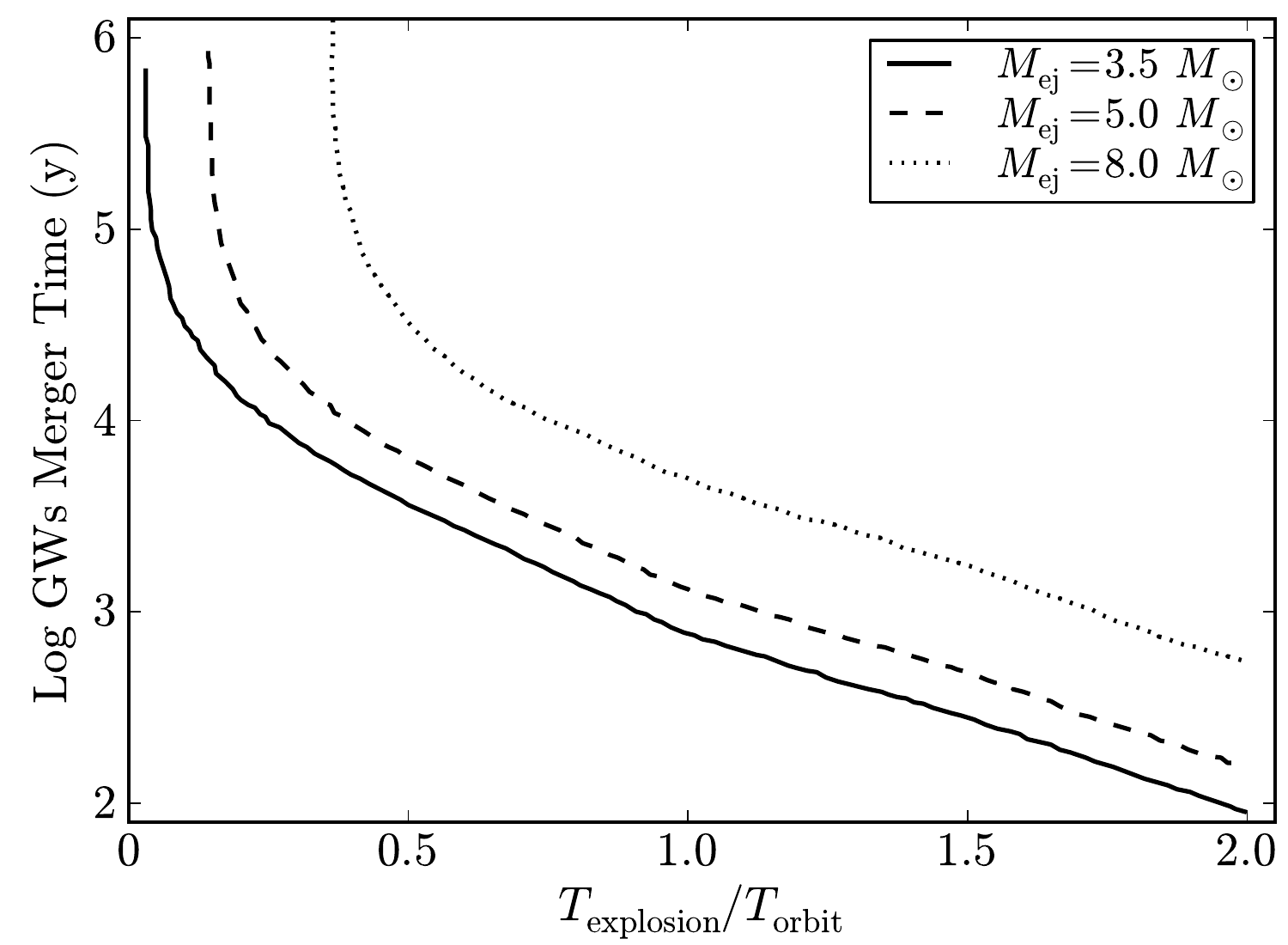}
    \caption{(\textbf{a}) Semi-major axis versus explosion time for three different mass ejecta scenarios: $3.5~M_\odot$ (solid), $5.0~M_\odot$ (dotted), $8.0~M_\odot$ (dashed), including mass accretion and momentum effects {(taken from Figure~2 in~\cite{2015PhRvL.115w1102F})}. Including these effects, all systems with explosion times above $0.7$ times the orbital time are bound and the final separations are on par with the initial separations. (\textbf{b}) Merger time due to gravitational wave emission as a~function of explosion time for the same three binaries of the left panel {(taken from Figure~3 in~\cite{2015PhRvL.115w1102F})}. Note that systems with explosion times $0.1$--$0.6~T_{\rm orbit}$ have merger times less than roughly $10^4$~y. For most of our systems, the explosion time is above this limit and we expect most of these systems to merge quickly.}
    \label{fig:PRL2015}
\end{figure}

We recall the specific examples studied in~\cite{2015PhRvL.115w1102F}: close binaries in an~initial circular orbit of radius $7\times10^9$~cm, CO$_{\rm core}$ radii of $(1$--$4) \times 10^9$~cm with a~2.0~$M_\odot$ NS companion. The CO$_{\rm core}$ leaves a~central 1.5~$M_\odot$ NS, ejecting the rest of the core. The NS leads to a~BH with a~mass equal to the NS critical mass. For these parameters it was there obtained that even if 70\% of the mass is lost the binary remains bound, providing the explosion time is of the order of the orbital period ($P=180$~s) with semi-major axes of less than $10^{11}$~cm (see~Figure~\ref{fig:PRL2015}).

The tight $\nu$NS-BH binaries produced by BdHNe will, in due time, merge owing to the emission of gravitational waves. For the above typical parameters the merger time is of the order of $10^4$~year, or~even less (see~Figure~\ref{fig:PRL2015}). We expect little baryonic contamination around such merger site since this region has been cleaned-up by the BdHN. These conditions lead to a~new family of sources which we have called ultrashort GRBs, U-GRBs.

%%%%%%%%%%%%%%%%%%%%%%%%%%%%%%%%%%%%%%%%%%%%%%%%%%%%%%%%%%%%
%%%%%%%%%%%%%%%%%%%%%%%%%%%%%%%%%%%%%%%%%%%%%%%%%%%%%%%%%%%%
\section{BdHN Formation, Occurrence Rate and Connection with Short GRBs}\label{sec:7}
%%%%%%%%%%%%%%%%%%%%%%%%%%%%%%%%%%%%%%%%%%%%%%%%%%%%%%%%%%%%
%%%%%%%%%%%%%%%%%%%%%%%%%%%%%%%%%%%%%%%%%%%%%%%%%%%%%%%%%%%%

%%%%%%%%%%%%%%%%%%%%%%%%%%%%%%%%%%%%%%%%%%%%%%%%%%%%%%%%%%%%
\subsection{An Evolutionary Scenario}
%%%%%%%%%%%%%%%%%%%%%%%%%%%%%%%%%%%%%%%%%%%%%%%%%%%%%%%%%%%%

{The X-ray binary and SN} communities have introduced a~new evolutionary scenario for the formation of compact-object binaries (NS-NS or NS-BH). After the collapse of the primary star forming a~NS, the binary undergoes mass-transfer episodes finally leading to the ejection of both the hydrogen and helium shells of the secondary star. These processes lead naturally to a~binary composed of a~CO$_{\rm core}$ and an~NS companion (see~Figure~\ref{fig:binaryevolution}). In the X-ray binary and SN communities these systems are called ``ultra-stripped'' binaries~\cite{2015MNRAS.451.2123T}. These systems are expected to comprise $0.1$--$1\%$ of the total SNe~\cite{2013ApJ...778L..23T}.

The existence of ultra-stripped binaries supports our scenario from the stellar evolution side. In~the above studies most of the binaries have orbital periods in the range $3\times 10^3$--$3\times 10^5$~s which are longer with respect to the short periods expected in the BdHN scenario. Clearly, XRF and BdHN progenitors should be only a~small subset that result from the binaries with initial orbital separation and component masses leading to CO$_{\rm core}$-NS binaries with short orbital periods, e.g.,~$100$--$1000$~s for the occurrence of BdHNe. This requires fine-tuning both of the CO$_{\rm core}$ mass and the binary orbit. From~an astrophysical point of view the IGC scenario is characterized by the BH formation induced by the hypercritical accretion onto the NS companion and the associated GRB emission. Indeed, GRBs are a~rare phenomenon and the number of systems approaching the conditions for their occurrence must be low (see~\cite{2015PhRvL.115w1102F} for details).

%%%%%%%%%%%%%%%%%%%%%%%%%%%%%%%%%%%%%%%%%%%%%%%%%%%%%%%%%%%%
\subsection{Occurrence Rate}
%%%%%%%%%%%%%%%%%%%%%%%%%%%%%%%%%%%%%%%%%%%%%%%%%%%%%%%%%%%%

If we assume that XRFs and BdHNe can be final stages of ultra-stripped binaries, then~the percentage of the ultra-stripped population leading to these long GRBs must be very small. The~observed occurrence rate of XRFs and BdHNe has been estimated to be $\sim$100~Gpc$^{-3}$~yr$^{-1}$ and $\sim$1~Gpc$^{-3}$~yr$^{-1}$, 
respectively~\cite{2016ApJ...832..136R}, namely the $0.5\%$ and $0.005\%$ of the Ibc SNe rate, \mbox{$2\times 10^4$}~Gpc$^{-3}$~yr$^{-1}$~\cite{2007ApJ...657L..73G}. It has been estimated that $(0.1$--$1\%)$ of the SN Ibc could originate from ultra-stripped binaries~\cite{2013ApJ...778L..23T}, which would lead to an~approximate density rate of $(20$--$200)$~Gpc$^{-3}$~yr$^{-1}$. This would imply that a~small fraction ($\lesssim$5\%) of the ultra-stripped population would be needed to explain the BdHNe while, roughly speaking, almost the whole population would be needed to explain the XRFs (see Table~\ref{tab:rates}). These numbers, while waiting for a~confirmation by further population synthesis analyses, would suggest that most SNe originated from ultra-stripped binaries should be accompanied by an~XRF. It is interesting that the above estimates are consistent with traditional estimates that only $\sim$0.001--1\% of massive binaries lead to double compact-object binaries~\cite{1999ApJ...526..152F,2012ApJ...759...52D,2014LRR....17....3P}.

%%%%%%%%%%%%%%%%%%%%%%%%%%%%%%%%%%%%%%%%%%%%%%%%%%%%%%%%%%%%
\subsection{Connection with Short GRBs}
%%%%%%%%%%%%%%%%%%%%%%%%%%%%%%%%%%%%%%%%%%%%%%%%%%%%%%%%%%%%

It is then clear that XRFs and BdHNe lead to $\nu$NS-NS and $\nu$NS-BH binaries. In due time, the~emission of gravitational waves shrink their orbit leading to mergers potentially detectable as short GRBs. This implies a~connection between the rate of long and short GRBs. It is clear from the derived rates (see Table~\ref{tab:rates} and~\cite{2016ApJ...832..136R,2018ApJ...859...30R}) that the short GRB population is dominated by the low-luminosity class of short Gamma-ray flashes (S-GRFs), double NS mergers that do not lead to BH formation. It can be seen that it is sufficient $\lesssim$4\% of XRFs to explain the S-GRFs population, which would be consistent with the fact that many XRF progenitor binaries will get disrupted by the SN explosion. Therefore, by now, the observed rates of the GRB subclasses are consistent with the interesting possibility of a~connection between the progenitors of the long and the ones of the short GRBs. 

%%%%%%%%%%%%%%%%%%%%%%%%%%%%%%%%%%%%%%%%%%%%%%%%%%%%%%%%%%%%
%%%%%%%%%%%%%%%%%%%%%%%%%%%%%%%%%%%%%%%%%%%%%%%%%%%%%%%%%%%%
\section{Conclusions}
%%%%%%%%%%%%%%%%%%%%%%%%%%%%%%%%%%%%%%%%%%%%%%%%%%%%%%%%%%%%
%%%%%%%%%%%%%%%%%%%%%%%%%%%%%%%%%%%%%%%%%%%%%%%%%%%%%%%%%%%%

It is by now clear that short and long Gamma-ray bursts subclassify into eight different families and have as progenitors binary systems of a~variety of flavors (see Table~\ref{tab:rates}). We have focused in this work on the specific class of BdHNe of two types: type I and type II BdHNe, what in our old classification~\cite{2016ApJ...832..136R} we called BdHNe and XRFs, respectively.

{We have devoted this article mostly to the theoretical aspects of the \emph{induced gravitational collapse scenario} and its evolution into BdHN as a~complete model of long GRBs. We have also discussed, although briefly, the observable features of the model and how they compare with the observational data, providing to the reader the appropriate references for deepening this important aspect.}

BdHNe I and II have as a~common progenitor a~CO$_\textrm{core}$-NS binary. The CO$_\textrm{core}$ explodes as type Ic SN, forming at its center a~new NS, which we denote $\nu$NS, and produces onto the NS companion a~hypercritical accretion process accompanied by an~intense neutrino emission. The intensity of the accretion process and the neutrino emission depends mainly on the binary period, being more intense for tighter binaries. The NS companion in such an~accretion process can reach or not the critical mass for gravitational collapse, i.e., to form a~BH. The former binaries leading to a~BH by accretion are the BdHNe I, while the ones in which the NS companion becomes just a~more massive NS, are the BdHNe II (the old XRFs) (see Table~\ref{tab:rates}).

We have reviewed the results of the numerical simulations performed of the above physical process starting from the 1D ones all the way to the latest 3D SPH ones. The simulation of this binary process has opened our eyes to a~new reality: long GRBS are much richer and more complex systems than every one of us thought before, with the 3D morphology of the SN ejecta, that becomes asymmetric by the accretion process, playing a~fundamental role in the GRB analysis. 

We have recalled the relevance of each of the following processes in a~BdHN: 
\begin{enumerate} [leftmargin=2.3em,labelsep=4mm]
\item[(1)] the SN explosion; 

\item[(2)] the hypercritical accretion onto the NS companion;
\item[(3)] the NS collapse with consequent BH formation; 

\item[(4)] the initiation of the \emph{inner engine};

\item[(5)] the $e^+e^-$ plasma production; 

\item[(6)] the $e^+e^-$ plasma feedback onto the SN which converts the SN into a~HN; 

\item[(7)] the formation of the cavity around the newborn BH; 

\item[(8)] the transparency of the $e^+e^-$ plasma along different directions; 

\item[(9)] the HN emission powered by the $\nu$NS; 

\item[(10)] the action of the \emph{inner engine} in accelerating protons leading to UHECRs and to the high-energy~emission.
\end{enumerate}

{The aforementioned involved physical processes in a~BdHN have specific signatures observable (and indeed observed) in the long GRB multiwavelength lightcurves and spectra. We have recalled for each process its energetics, spectrum, and associated Lorentz factor: from the mildly-relativistic X-ray precursor, to the ultrarelativistic prompt Gamma-ray emission, to the mildly-relativistic X-ray flares of the early afterglow, to the mildly-relativistic late afterglow and to the high-energy GeV emission.}

All of the above is clearly in contrast with a~simple GRB model attempting to explain the entire GRB process with the kinetic energy of an~ultrarelativistic jet extending through all of the above GRB phases, as in the traditional collapsar-fireball model. 

If the binaries keep bound during the explosion, BdHNe I lead to $\nu$NS-BH binaries and BdHNe II lead to $\nu$NS-NS binaries. In due time, via gravitational wave emission, such binaries merge producing short GRBs. This unveiled clear interconnection between long and short GRBs and their occurrence rates needs to be accounted for in the cosmological evolution of binaries within population synthesis models for the formation of compact-object binaries.

{We have taken the opportunity to include a~brief summary of very recent developments published during the peer-review process of this article. These results cover the explanation of the observed GeV emission in BdHNe~\cite{2018arXiv181101839R,2018arXiv181200354R,2019arXiv190404162R,2019arXiv190403163R}. One of the most relevant aspects of this topic is that it requests the solution of one of the fundamental problems in relativistic astrophysics: how to extract the rotational energy from a~BH. This implies the role of a~magnetic field around the newborn BH and the presence of surrounding matter as predicted in a~BdHN. We have called this part of the system the \emph{inner engine} of the high-energy emission. The BH rotation and surrounding magnetic field, for appropriate values induces an~electric field via the Wald's mechanism~\cite{1974PhRvD..10.1680W}. Such an~electric field is of paramount importance in accelerating surrounding protons to ultrarelativistic velocities leading to the high-energy emission via proton-synchrotron radiation. The details of this exciting new topic are beyond the scope of the present article but we encourage the reader to go through the above references for complementary details.}

\vspace{6pt}

%\authorcontributions{\hl{For research articles with} several authors, a~short paragraph specifying their individual contributions must be provided. The following statements should be used ``conceptualization, X.X. and Y.Y.; methodology, X.X.; software, X.X.; validation, X.X., Y.Y. and Z.Z.; formal analysis, X.X.; investigation, X.X.; resources, X.X.; data curation, X.X.; writing--original draft preparation, X.X.; writing--review and editing, X.X.; visualization, X.X.; supervision, X.X.; project administration, X.X.; funding acquisition, Y.Y.'', please turn to the  \href{http://img.mdpi.org/data/contributor-role-instruction.pdf}{CRediT taxonomy} for the term explanation. Authorship must be limited to those who have contributed substantially to the work reported.}

%%%%%%%%%%%%%%%%%%%%%%%%%%%%%%%%%%%%%%%%%%
\funding{This research received no external funding.
%``This research was funded by NAME OF FUNDER grant number XXX.'' and  and ``The APC was funded by XXX''. Check carefully that the details given are accurate and use the standard spelling of funding agency names at \url{https://search.crossref.org/funding}, any errors may affect your future funding.
}

%%%%%%%%%%%%%%%%%%%%%%%%%%%%%%%%%%%%%%%%%%
%\acknowledgments{In this section you can acknowledge any support given which is not covered by the author contribution or funding sections. This may include administrative and technical support, or donations in kind (e.g., materials used for experiments).}

%%%%%%%%%%%%%%%%%%%%%%%%%%%%%%%%%%%%%%%%%%
\conflictsofinterest{The authors declare no conflict of interest. 
%Authors must identify and declare any personal circumstances or interest that may be perceived as inappropriately influencing the representation or interpretation of reported research results. Any role of the funders in the design of the study; in the collection, analyses or interpretation of data; in the writing of the manuscript, or in the decision to publish the results must be declared in this section. If there is no role, please state ``The funders had no role in the design of the study; in the collection, analyses, or interpretation of data; in the writing of the manuscript, or in the decision to publish the results''.
}

\reftitle{References}

%%%%%%%%%%%%%%%%%%%%%%%%%%%%%%%%%%%%%%%%%%
\end{document}